\documentclass[twocolumn,tighten,numberedappendix]{aastex631}

\usepackage{amsmath}
\usepackage{ulem}
\usepackage{stmaryrd}
\setcitestyle{notesep={ }}

\usepackage{graphicx}
\usepackage{dcolumn}
\usepackage{bm}
\usepackage{hyperref}

\usepackage[T1]{fontenc}

\usepackage{scalerel,tikz}

\usepackage{fancyvrb}
\newcommand{\mbf}[1]{\bm{#1}}
\newcommand{\del}{\nabla}
\usepackage[flushleft]{threeparttablex}
\usepackage{cprotect}
\usepackage{hyperref}

\defcitealias{2023ApJ...959...34B}{B23}

\begin{document}




\title{Global Magnetohydrodynamic Simulations of Monster Shocks in Neutron Star Magnetospheres}


\correspondingauthor{Michael P. Grehan}
\email{michael.grehan@mail.utoronto.ca}

\author[0009-0003-1842-192X]{Michael P. Grehan} 
\affil{Department of Physics, University of Toronto, 60 St. George Street, Toronto, ON M5S 1A7, Canada}
\affil{Canadian Institute for Theoretical Astrophysics, 60 St. George Street, Toronto, ON M5S 3H8, Canada}

\author[0000-0002-7301-3908]{Bart Ripperda}
\affil{Canadian Institute for Theoretical Astrophysics, 60 St. George Street, Toronto, ON M5S 3H8, Canada}
\affil{Department of Physics, University of Toronto, 60 St. George Street, Toronto, ON M5S 1A7, Canada}
\affil{D. A. Dunlap Department of Astronomy, University of Toronto, Toronto, ON M5S 3H4, Canada}
\affil{Perimeter Institute for Theoretical Physics, Waterloo, ON N2L 2Y5, Canada}

\author[0000-0001-5660-3175]{Andrei M. Beloborodov} 
\affil{Physics Department and Columbia Astrophysics Laboratory, Columbia University, 538 West 120th Street New York, NY 10027,USA}
\affil{Max Planck Institute for Astrophysics, Karl-Schwarzschild-Str. 1, D-85741, Garching, Germany}

\author[0000-0003-4305-5653]{Christopher Thompson}
\affil{Canadian Institute for Theoretical Astrophysics, 60 St. George Street, Toronto, ON M5S 3H8, Canada}

\author[0000-0002-0491-1210]{Elias R. Most} 
\affil{TAPIR, Mailcode 350-17, California Institute of Technology, 1200 E California Blvd, Pasadena, CA 91125, USA}
\affil{Walter Burke Institute for Theoretical Physics, California Institute of Technology, Pasadena, CA 91125, USA}


\begin{abstract}
    Waves launched from the neutron star surface or inner magnetosphere propagate through the magnetosphere as small perturbations, but can grow relative to the background magnetic field and steepen into ``monster shocks'' --- ultra-relativistic magnetized shocks which can power high-energy emission. Such shocks can develop around isolated magnetars, merging  binaries, and collapsing neutron stars. They occur in magnetically dominated plasma and are described by relativistic magnetohydrodynamics (MHD). We present global relativistic MHD simulations of monster shocks in unperturbed and perturbed (``wrinkled'') backgrounds with a global dipolar geometry. Our simulations confirm analytical predictions for equatorial shocks and provide new insight into the behavior of oblique shocks off the equator. Simulations where the shock is formed through Alfv\'{e}n mode to  fast mode conversion are also presented, demonstrating the generic nature of the monster shock mechanism. We explore how the presence of additional modes in the magnetosphere modifies the shock behavior. Modes of comparable amplitude can fragment the shock front, substantially reduce the magnetization, produce localized enhancements in the Lorentz factor relative to an unperturbed dipole background, and intermittently generate additional shocks along a line of sight. 
\end{abstract}


\section{Introduction}\label{sec:intro}


The bright outbursts of magnetars are powered by rapid changes in their ultrastrong magnetic fields, which typically exceed the Schwinger field, $B_{\rm{Q}} = 4.4 \times 10^{13}$ G \citep{1992ApJ...392L...9D}.   Electromagnetic radiation of high intensity is occasionally emitted in the form of fast radio bursts (FRBs; 
\citealt{CHIMEFRB:2020abu,Bochenek2020}), X-ray bursts, and even more
luminous magnetar giant flares \citep{2017ARA&A..55..261K}.
The galactic magnetar SGR 1935+2154 has twice been observed to produce FRBs coinciding with an X-ray burst \citep{2020ApJ...898L..29M,2026arXiv260210895W}.
Transient emission is powered by dissipation of the enormous magnetic energy reservoir through a variety of channels (e.g., current driven instabilities, Alfv\'{e}nic cascades, shocks), which are thought to be triggered by internal motion and stress within the star.

A promising mechanism leading to rapid magnetic dissipation is the monster shock 
(\citealt{2023ApJ...959...34B}, hereafter \citetalias{2023ApJ...959...34B}; \citealt{2022arXiv221013506C}).
When a fast magnetosonic (FMS) wave is launched into the 
magnetar magnetosphere it steepens into an ultra-relativistic shock.
At small amplitude, $\delta B/B \ll 1$, the wave propagates through the highly magnetized background as a vacuum wave, with wave field decaying inversely with radius,
$\delta B \propto 1/r$.
The amplitude relative to a background dipole field $B$ steepens as $\delta B/B \propto r^2$,
creating zones where the wave and background magnetic fields partly cancel and $E^2\rightarrow B^2$.  
This signals the breakdown of a force-free description and the dissipation of energy into particles
\citep{2012PhRvD..86j4035L}. Here we deploy two-dimensional relativistic magnetohydrodynamic (MHD) simulations
with relativistic magnetization to test and demonstrate the process of monster shock formation:  how the shock Lorentz
factor scales with magnetization, the efficient dissipation of fast wave energy, and small-scale structure 
in the shock imparted by pre-existing perturbations to the stellar magnetic field.

The external magnetic field
is tied to the outer layers of the magnetar and must respond to motions of the stellar core and solid crust.
A fraction of the energy deposited in the external field may be converted to fast waves by a variety of
mechanisms.  The most efficient mechanism described so far involves exciting
dynamic twist 
around the poles \citep{2021ApJ...908..176Y,2024ApJ...972..139M, 2025ApJ...980..222B};
a few percent of the outburst energy may also be converted directly to fast waves when the crust
experiences a localized quake \citep{2026ApJ...998..190Q,2025ApJ...995L..57B}.
Fast waves of even higher energy may form during and after the merger of a neutron star with another compact star: through a collision between two magnetospheres \citep{2025ApJ...981L..17M},
during the collapse of a hypermassive magnetar \citep{2012PhRvD..86j4035L, 2024ApJ...974L..12M}, and in the related case of a non-disruptive merger of a black hole -- neutron star binary \citep{2025ApJ...982L..54K}.  Fast wave emission therefore typically takes place in the presence of overlapping waves,
including current-carrying Alfv\'en (A) waves; 
this motivates a study of how shock formation may be perturbed by non-linear wave interactions.


Dissipation from one or more monster shocks will contribute to the X-ray emission during a magnetar outburst, possibly coinciding with FRBs.  The shock can also be a direct source of low frequency
precursor radiation through the maser instability \citep{1988PhFl...31..839A, 1991PhFlB...3..818H, 2021PhRvL.127c5101S}.  One-dimensional kinetic simulations
suggest that up to $10^{-3}$ of the shock energy may convert to radio waves for a certain range of background plasma density, as the emission frequency depends on the upstream plasma skin depth \citep{2025PhRvL.134c5201V}. The generation of precursor emission is a kinetic effect and therefore not captured in the MHD simulations presented in this paper. 

To date, there have been few simulations showing the formation of a monster shock in a realistic magnetic background.
One-dimensional (1D) kinetic simulations by \cite{2022arXiv221013506C} and \cite{2025PhRvL.134c5201V} allow for expanded scale separation and resolution, enabling the precursor mechanism to be resolved. However, the shock mechanism remains in the MHD regime and these 1D simulations do not include a realistic dipolar background magnetic field.
Global MHD simulations of neutron star collapse and binary merger in dynamical spacetimes \citep{2024ApJ...974L..12M, 2025ApJ...982L..54K} have also demonstrated
shock formation, albeit with limited magnetization and resolution.   Most recently, global kinetic simulations of monster shocks from pure fast modes propagating in an
unperturbed dipole field have been performed \citep{2025arXiv250604175B}. The results match analytic predictions \citepalias{2023ApJ...959...34B} and demonstrate the precursor
effect. Here, our MHD simulations, by construction, neglect any kinetic effects and therefore the impact of scale separation on the MHD shock mechanism is of no concern. Force-free electrodynamic simulations reveal fast wave propagation and steepening but, by construction, are unable to describe shock formation and dissipation
\citep{2024ApJ...972..139M, 2025ApJ...980..222B,2025ApJ...995L..57B}.

Crucially, theoretical analysis and simulations have focused primarily on the equatorial perpendicular shock, or at most on a perfectly dipolar magnetosphere. Magnetars are expected to have twisted magnetospheres due to internal stresses \citep{Thompson:2001ig, 2007ApJ...657..967B, 2017ARA&A..55..261K}, along with the presence of other modes in the magnetosphere \citep{2026ApJ...998..190Q, 2025ApJ...995L..57B}. Similarly, magnetospheres of merging or collapsing stars will be perturbed from a pure dipole and be filled with additional waves \citep{2012PhRvD..86j4035L, 2024ApJ...974L..12M, 2025ApJ...981L..17M, 2025ApJ...982L..54K}. We perform the first MHD
simulations of non-linear wave launching (by A wave collisions) and of shock formation in the presence
of a wrinkled dipole magnetic field supporting secondary waves. Our novel MHD method allows for the high magnetization required to achieve the asymptotic monster shock regime, where the upstream Lorentz factor greatly exceeds unity and the FMS wavelength is much smaller than the non-linear radius, as well as the presence of a dissipation mechanism to kinetic and thermal plasma energy.

\begin{figure*}
    \centering
    \includegraphics[width=\linewidth]{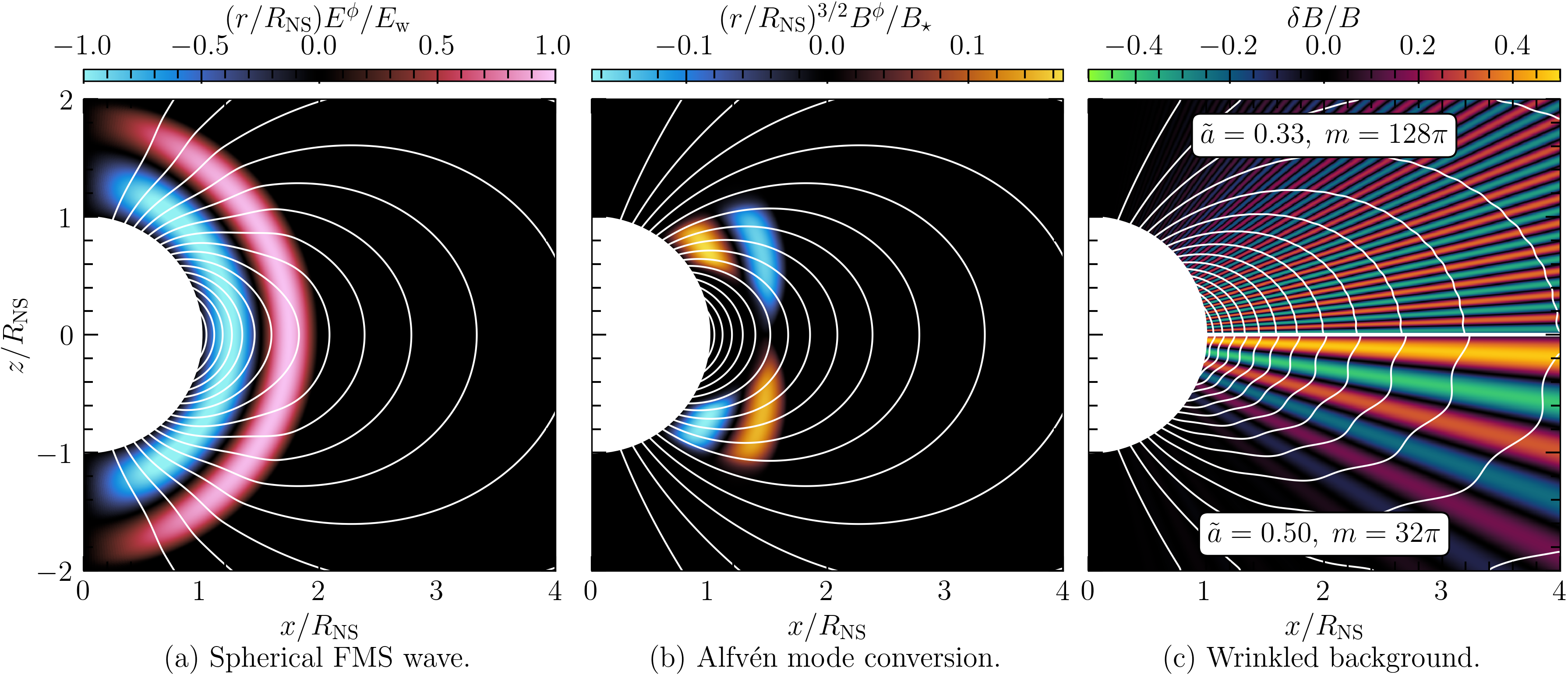}
    \caption{The three setups presented in this paper. Left: a spherical FMS wave, $E^\phi = R_{\rm NS} E_{\rm w} \sin(\omega t)/r$, is launched from the neutron star surface into a dipolar background magnetic field. Middle: toroidal A waves, $B^\phi \propto (R_{\rm NS}/r)^{3/2}$, of opposite polarity on either side of the equator are launched through a localized twist on the stellar surface (Equation~\eqref{eq:twist}). The waves collide at the equator producing a fast mode which propagates radially outwards and shocks. Right: the initial background dipolar magnetic field is perturbed with a harmonic standing wave (Equation~\eqref{eq:wiggles}), resulting in a wrinkled background. A fast mode is then launched from the stellar surface and interacts with this wave perturbation as it begins to shock. The ratio of wrinkle and background field, $\delta B/B$, is shown for two values of the wrinkle amplitude $\tilde{a}$ and wave number $m$ (Equation~\eqref{eq:wiggles}). White lines show linearly spaced magnetic flux contours.}
    \label{fig:three_subfigs}
\end{figure*}

\subsection{Plan of the Paper}

The paper is organized as follows.   Section~\ref{sec:initial_cons} describes the initial conditions for the three families of models
investigated here:
(1) a spherical 
FMS wave directly launched into a dipolar magnetosphere from the stellar boundary; 
(2) Alfv\'{e}n waves launched through localized twists on the stellar boundary, which interact to form a compressive wave; and
(3) a spherical fast mode launched through a dipolar magnetosphere that is perturbed by 
higher-frequency modes around the shock formation radius. 
In Section~\ref{sec:results1}, we describe the results of simulations (1) 
and compare them with theoretical predictions and recent
global kinetic simulations.
Section~\ref{sec:pertresults} describes the results of simulations (2) and (3), in particular the qualitative changes to shock structure
and entropy profile that may be imparted by a wrinkled background.
Our results are summarized in
Section~\ref{sec:conclusion}, 
where we discuss the relevance to astrophysical bursts from magnetars, neutron star mergers and collapse. 

Throughout this paper, the numerical value of quantity $X$ is normalized by $X = X_n \times  10^n$ in cgs units.

\newpage

\section{Method and Simulation Setups}\label{sec:initial_cons} 

We consider wave propagation in a dipolar magnetic field surrounding a spherical star.
Electrically conducting plasma embedded in this magnetic field is given the equation of state of 
a relativistic ideal gas with adiabatic index $\hat\gamma = 4/3$;  the medium outside the star 
is evolved as an ideal magnetofluid with a large ratio of magnetic to plasma pressure.
The dipolar field is initiated in flat space spherical coordinates $(r,\theta,\phi)$ 
through the vector potential
\begin{equation}
     \mathbf{A}_{\rm{bg}} =  \mu \left( \frac{\sin\theta}{r^2} \right) \hat{\phi},
\end{equation}
where $\mu = B_\star R_{\rm NS}^3$ is the magnetic moment of the star, $B_\star$ is the surface magnetic field strength on the equator, $R_{\rm NS}$ is the stellar radius, and “bg” denotes the “background”. 

The surface of the star is treated as a perfect conductor, anchoring magnetic field lines into the stellar surface. 
At the outer boundary, zero-gradient conditions are applied to the fluid variables ($\bm{v}$, $\rho$, $p$). The tangential component of the magnetic field, $B^\theta$, is extrapolated assuming a dipolar radial scaling ($\propto r^{-3}$). The toroidal component, $B^\phi$, is assigned a zero-gradient condition, while the radial component, $B^r$, is adjusted to enforce the divergence-free constraint, $\bm{\nabla} \cdot \bm{B} = 0$.
Evolved quantities are treated as symmetric across the polar boundaries,
except for the perpendicular components of vector quantities, 
which are antisymmetric. The pressure and all components of the velocity are initiated as zero in the background.  The effects of stellar rotation are neglected, given that active
magnetars tend to rotate slowly, with periods $P \sim 1-10\,$s \citep{2017ARA&A..55..261K}.
As a result,
nonlinear wave interactions and wave breaking already occur well within the
corotating magnetosphere.  

\begin{table*}[htbp] 
\caption{\label{tab:sims}%
Spherical FMS Wave Parameters. 
} 
\begin{ruledtabular}
\begin{tabular}{ccc|cc|cc|ccc}

&\multicolumn{2}{c|}{\textbf{Varying Magnetization}} & \multicolumn{2}{c|}{\textbf{Varying Amplitude}}  & \multicolumn{2}{c|}{\textbf{Varying Wavelength}} &
\multicolumn{2}{c}{\textbf{Multiple Waves}} & \\

\hline

&
\multicolumn{1}{c}{\textrm{Sim. ID}}&
\multicolumn{1}{c|}{\textrm{$\sigma_\times$}}&

\multicolumn{1}{c}{\textrm{Sim. ID}}&
\multicolumn{1}{c|}{\textrm{$E_{\rm{w}}/B_\star$}}&

\multicolumn{1}{c}{\textrm{Sim. ID}}&
\multicolumn{1}{c|}{\textrm{$\lambda \, [R_\times]$}} & 

\multicolumn{1}{c}{\textrm{Sim. ID}}&
\multicolumn{1}{c}{\textrm{$N$}} &
\\
&
\multicolumn{1}{c}{(1)} & \multicolumn{1}{c|}{(2)} & 
\multicolumn{1}{c}{(1)} & \multicolumn{1}{c|}{(3)} &
\multicolumn{1}{c}{(1)} & \multicolumn{1}{c|}{(4)}  & 

\multicolumn{1}{c}{(1)} & \multicolumn{1}{c}{(5)} & 
\\
\hline
&
\texttt{sigma25} & 25 & \texttt{eW01} & 0.1 & \texttt{w2pi} & 0.45 & 
\texttt{mw1} & 1 & \\
&
\texttt{sigma50} & 50 & \texttt{eW005} & 0.05 & \texttt{w4pi} & 0.22 & 
\texttt{mw2} & 2 & \\
&
\texttt{sigma75} & 75 & \texttt{eW001} & 0.01 & \texttt{w8pi} & 0.11 & 
\texttt{mw3} & 3 & \\

&
\texttt{sigma100} & 100 & & & & & 
\texttt{mw4} & 4 & \\

&
&  & & & & & 
\texttt{mw3opp} & 3 & \\

\end{tabular}

\end{ruledtabular}
\begin{tablenotes}[para]
        \textit{\textbf{Notes.}} Simulation parameters for the initial conditions described in Section~\ref{sec:sphereicalwave}. All simulations have an effective resolution of $(16384, 2048)$ cells in $(r,\theta)$ with $r/R_{\rm{NS}}\in[1,10]$ and $\theta\in[0,\pi]$. The magnetization is uniform up to the non-linearity radius, after which it decays as $(R_\times/r)^3$ in all simulations but \texttt{mw3opp}, for which it is uniform. The FMS wave is launched with the $\bm{B}_{\rm w}\cdot\bm{B}_{\rm bg}<0$ portion leading in all simulations except for \texttt{mw3opp} which leads with $\bm{B}_{\rm w}\cdot\bm{B}_{\rm bg}>0$. 
        \textbf{Column (1):} the unique simulation ID. 
        \textbf{Column (2):} the magnetization, $\sigma_\times$ at $r=R_\times$ which then drops off as $(r/R_{\times})^{-3}$. Simulations without a stated magnetization have $\sigma_\times=50$.
        \textbf{Column (3):} the amplitude of the fast waves launched from the stellar surface relative to 
        the surface equatorial magnetic field.  Simulations without a stated wave amplitude have $E_{\rm{w
        }}/B_\star=0.1$.
        \textbf{Column (4):} the wavelength of the fast waves launched from the stellar surface in units of the non-linearity radius. Simulations without a stated wavelength have frequency $\omega = 2\pi c/\lambda = 2\pi \,  [c/R_{\rm{NS}}]$, except for \texttt{mw3opp} which has $\omega = 4\pi \,  [c/R_{\rm{NS}}]$.
        \textbf{Column (5):} the number of wavelengths launched from the boundary. Simulations without a stated value launch two wavelengths.
    \end{tablenotes}
\end{table*}

\subsection{Spherical 
Fast Magnetosonic Wave}\label{sec:sphereicalwave}

To study the direct injection of a fast wave into the magnetosphere, we set the electric field at the stellar surface to
\begin{equation}
    \bm{E}(r=R_{\rm{NS}}) = \bm{E}_{\rm{w}} = E_{\rm{w}} \sin(\omega t) \hat{\phi}, \label{eq:fms}
\end{equation}
where $\omega$ is the wave frequency and $E_{\rm{w}}$ is the surface wave amplitude. 
A tanh windowing function is used to set $\bm{E}=\bm{0}$ at the poles, in order to avoid numerical complications. 
In ideal MHD, the electric field is introduced by setting the fluid velocity to the
corresponding drift velocity,
\begin{equation}
    \bm{\beta}_D = \frac{\bm{E}_{\rm{w}}\times \bm{B} }{B^2} \simeq \frac{\bm{E}_{\rm{w}}\times \bm{B}_{\rm{bg}} }{B_{\rm{bg}}^2}.
\end{equation}
This results in the launching of a spherical 
electromagnetic wave. The fast mode amplitude scales as $E_{\rm{w}}, B_{\rm{w}}\propto r^{-1}$ 
and therefore $E_{\rm{w}}/B_{\rm{bg}} \propto r^2$. 
The propagation of the wave 
prior to shocking is shown in the left panel of Figure~\ref{fig:three_subfigs}.

The plasma in the magnetosphere is initiated with uniform magnetization $\sigma_\times$ within the radius where the equatorial 
wave reaches half the background magnetic field and
${\bm E}\times{\bm B}$ drift becomes relativistic, 
\begin{equation}
    R_\times = \frac{R_{\rm{NS}}}{\sqrt{2 E_{\rm{w}}/B_\star}}. \label{eq:Rx}
\end{equation}
In the default profile, the magnetization decays beyond this point, as 
\begin{equation}\label{eq:sigvsr}
\sigma_{\rm bg}(r) = \frac{B_{\rm{bg}}^2}{4\pi \rho_{\rm{bg}}c^2 } =
\sigma_\times
\begin{cases}
1, & r < R_\times, \\[6pt]
\left( \dfrac{R_\times}{r} \right)^3, & r \ge R_\times .
\end{cases}
\end{equation}
We choose a high magnetization
at the non-linearity radius, $\sigma_\times = 25-100$, thereby ensuring that the 
Lorentz factor upstream of the shock greatly exceeds unity.  Closer to the star, 
the wave propagation is nearly
force-free and depends weakly on $\sigma_{\rm bg}$ as long as the magnetization is large.
The tapering of $\sigma_{\rm bg}$ at large radius aids stability at high magnetization in 
the non-linear development of the electromagnetic pulse (see \citealt{2006ApJ...641..626N, 2018ApJ...859...71S, Ripperda:2019lsi, 2021PhRvD.103b3018K} for related considerations). A single simulation is performed with a uniform magnetization profile to probe shock evolution at large radii. 

Simulations using these initial conditions are listed in Table~\ref{tab:sims}, where we vary the magnetization, amplitude and frequency to verify the scalings obtained in \citetalias{2023ApJ...959...34B}. 
As well, we vary the number of waves launched to analyze the interaction between wavefronts.

        
        

        

\subsection{Alfv\'{e}n Mode Conversion}\label{sec:awcon}


Fast waves may be injected indirectly by exciting A waves of opposite polarity in opposing
hemispheres.
The conversion ${\rm A} + {\rm A} \rightarrow {\rm FMS}$ is strongest around the equator and most
efficient
when the A waves are in phase \citep{2024ApJ...972..139M}.
The A waves 
are launched by imposing a localized, axisymmetric twist on the stellar surface \citep{2013ApJ...774...92P},
\begin{equation}
    \bm{\omega} (\theta) = \frac{\omega_0 \sin(\omega_{\rm{A}} t)}{1+\exp\{ \kappa (|\theta - \theta_0|  - \Delta)\}} \hat{\phi},
    \label{eq:twist}
\end{equation}
where $\omega_0$ is the amplitude of the twist, $\theta_0$ is the center of the twist, $\Delta$ is the angular half-width, and $\omega_{\rm{A}}$ is the frequency of the twist.
If their amplitude is not too large, such that they remain linear while propagating along the dipolar field lines, the A waves are contained on the closed magnetic flux tube upon which they are launched and their interactions produce FMS waves \citep{2022ApJ...933..174Y, 2024ApJ...972..139M, 2025ApJ...980..222B}.
We do not consider the case of large amplitude twists where the excited A waves open the magnetosphere \citep{2022ApJ...933..174Y}.
The propagation of the toroidal A waves along the dipolar field prior to colliding 
is shown in the middle panel of Figure~\ref{fig:three_subfigs}.
In this set of models, we adopt a
uniform background magnetization $\sigma_{\rm bg}$.
The four chosen sets of initial conditions are listed in Table~\ref{tab:sims_AWAW};
these vary the amplitude and frequency of the localized surface twist and the excited A waves.

\begin{table}[htbp] 
\caption{\label{tab:sims_AWAW}%
Alfv\'{e}n Mode Conversion Parameters.} 
\begin{ruledtabular}
\begin{tabular}{ccc}

Sim. ID & $\omega_0\, [c/R_{\rm{NS}}]$ & $\omega_{\rm{A}} [c/R_{\rm{NS}}]$ \\
(1) & (2) & (3) \\

\hline


\texttt{AW022pi} & 0.2 & $2\pi$ \\

\texttt{AW032pi} & 0.3 & $2\pi$ \\
\texttt{AW034pi} & 0.3 & $4\pi$ \\
\texttt{AW038pi} & 0.3 & $8\pi$ \\


\end{tabular}

\end{ruledtabular}
\begin{tablenotes}[para]
        \textit{\textbf{Notes.}}  Simulation parameters for the initial conditions described in Section~\ref{sec:awcon}. All simulations have an effective resolution of $(16384, 4096)$ cells in $(r,\theta)$ with $r/R_{\rm{NS}}\in[1,10]$ and $\theta \in [0,\pi]$. All simulations launch two wavelengths of the torsional 
        wave mirrored across the equator with  $\kappa = 50$, $\Delta = 0.05 \pi$, $\theta_0 =  \pi/4$ (Equation~\eqref{eq:twist}), and uniform magnetization $\sigma_{\rm{bg}}=100$. 
        \textbf{Column (1):} the unique simulation ID. 
        \textbf{Column (2):} the amplitude of the localized twist applied at the stellar boundary in units of the inverse stellar crossing times.
        \textbf{Column (3):} the frequency of the localized twist in units of the inverse stellar crossing times.
    \end{tablenotes}
\end{table}

\subsection{Wrinkled Background}\label{sec:wiggles}

The interaction of the FMS wave with other modes in the magnetosphere could modify the formation of the monster shock or potentially suppress it through dissipative interactions. Recent three-dimensional force-free electrodynamic simulations \citep{2025ApJ...995L..57B}, coupled to a time-dependent magnetoelastic model of the crust \citep{2026ApJ...998..190Q}, have revealed a variety of nonlinear interactions.  Multiple compressive modes are excited that carry away energy from the magnetosphere and sometimes form caustics.  
Although these force-free simulations cannot capture shock dissipation, they help to motivate further
study of FMS wave propagation in a perturbed magnetosphere.  

To this end, we add a simple harmonic term to the dipole vector potential, 
\begin{equation}
    \bm{A} = \bm{A}_{\rm{bg}} + \bm{A}^\prime,
\end{equation}
where
\begin{equation}
    \bm{A}^\prime = \tilde{a} \left( \frac{\pi}{m} \right) \cos(m \theta/\pi ) \sin(\theta) \bm{A}_{\rm{bg}}, \label{eq:wiggles}
\end{equation}
$\tilde{a}$ is a dimensionless amplitude, and $m$ an integer. 
This initial condition is a superposition of oppositely propagating modes with a spatially dependent frequency
$\omega' \sim m/\pi r$.  The time-evolved ``wrinkles'' will spontaneously develop radial shear 
through linear phase mixing.
The right panel of Figure~\ref{fig:three_subfigs} shows the relative wrinkle amplitude and perturbed
poloidal magnetic field lines 
for two choices of $\tilde{a}$ and $m$.

The wrinkle perturbation introduces distinct zones of constructive and destructive interference
with the FMS wave.  It can induce fragmentation of a forming shock and allows us to investigate the 
appearance of secondary shocks along a given line of sight.
We do not simulate the interaction with non-axisymmetric A waves, whose effect on a low-frequency
FMS waves has been investigated in the quasi-linear, force-free regime \citep{golbraikh2023}.

 The radial profile of $\sigma_{\rm bg}$ follows Equation~\eqref{eq:sigvsr} in this set of models.
We tested a range of wrinkle amplitudes and wavenumbers in six simulations,
as summarized in Table~\ref{tab:sims_wiggles}.

\begin{table}[htbp] 
\caption{\label{tab:sims_wiggles}%
Wrinkled Background Parameters.} 
\begin{ruledtabular}
\begin{tabular}{ccc}

Sim. ID & $\tilde{a}$ & $m$ \\
(1) & (2) & (3) \\

\hline

\texttt{Wig0364} & 0.3 & 64$\pi$ \\
\texttt{Wig0564} & 0.5 & 64$\pi$ \\
\texttt{Wig0332} & 0.3 & 32$\pi$ \\
\texttt{Wig0532} & 0.5 & 32$\pi$ \\
\texttt{WigNoFMS} & 0.5 & 32$\pi$ \\
\texttt{NoWig} & 0.0 & --- \\

\end{tabular}

\end{ruledtabular}
\begin{tablenotes}[para]
        \textit{\textbf{Notes.}}  Simulation parameters for the initial conditions described in Section~\ref{sec:wiggles}. All simulations have an effective resolution of $(8192,8192)$ cells in $(r, \theta)$ with $r/R_{\rm{NS}}\in[1,10]$ and $\theta\in[0,\pi]$. All simulations launch a single spherical FMS wave with $\omega = 2\pi \, [c/R_{\rm{NS}}]$, $E_{\rm{W}}/B_\star=0.1$, and $\sigma_\times =100$, except for \texttt{WigNoFMS} which launches no FMS, to benchmark the effect of the standing waves.
        \textbf{Column (1):} the unique simulation ID. 
        \textbf{Column (2):} the maximum amplitude of the perturbation with respect to the background dipolar field.
        \textbf{Column (3):} the wavenumber of the perturbation.
    \end{tablenotes}
\end{table}

\subsection{Numerical Methods}

All simulations in this paper adopt 
the relativistic ideal MHD framework and use the Black Hole Accretion Code (\Verb+BHAC+) \citep{Porth:2016rfi, Olivares2019}.
Further details of the numerical model are presented in Appendix~\ref{app:num_methods}.
Details of the flooring, adaptive mesh refinement (AMR), and settings for numerical diffusion employed in this paper are presented in Appendix~\ref{app:amr_flooring}. Convergence tests are presented in Appendix~\ref{app:convergence}, where we 
verify that the strongest shock and smallest wavelength used in the spherical
FMS wave configuration 
(Table~\ref{tab:sims}) are well resolved.

\begin{figure*}
    \centering
    \includegraphics[width=\textwidth]{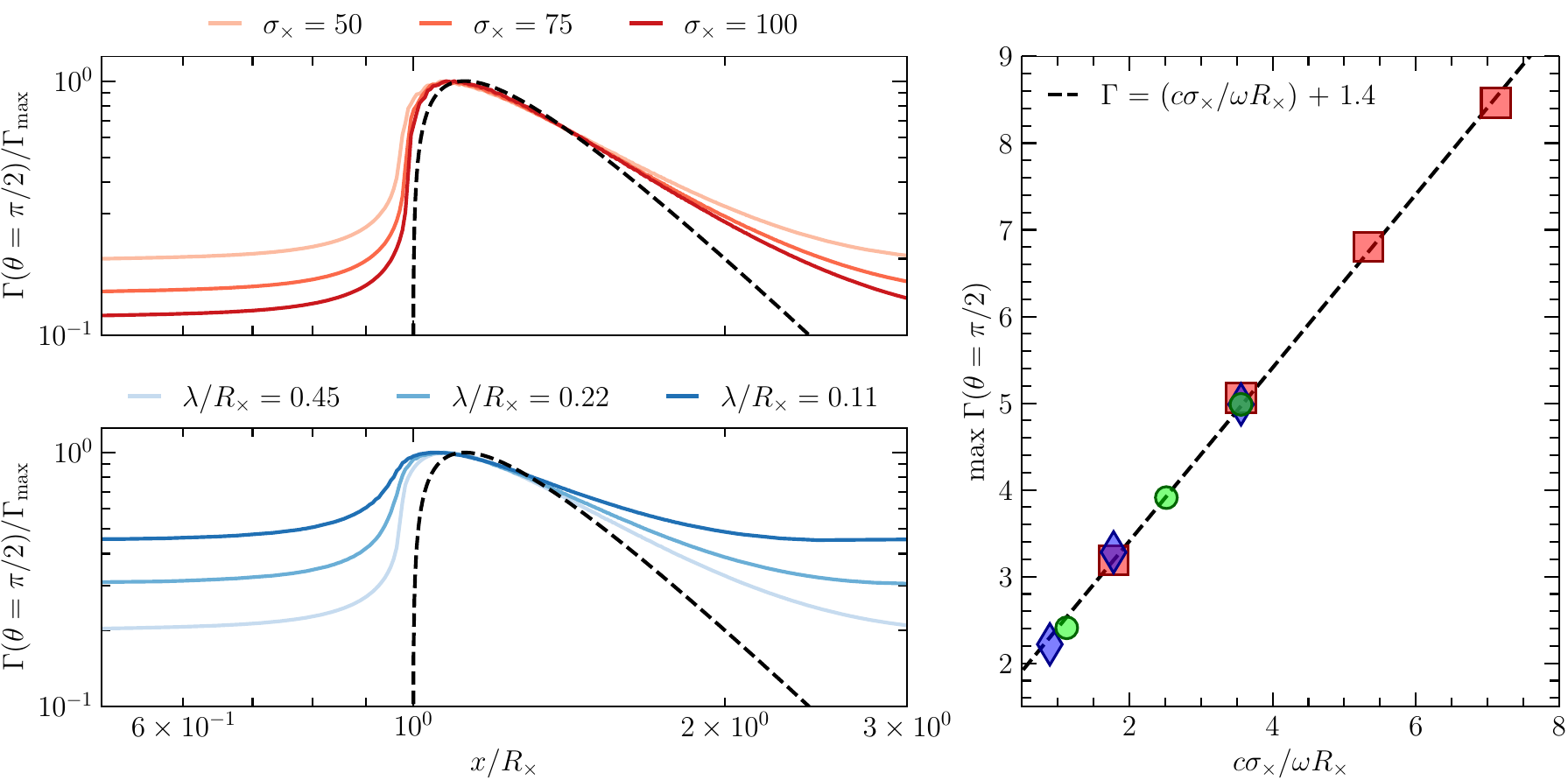}
\caption{
Comparison of the equatorial monster shock to analytic predictions \citepalias{2023ApJ...959...34B}.
Left: The upstream Lorentz factor of the monster shock on the equator as a function of cylindrical radius, plotted for multiple values of $\sigma_\times$ and $\lambda=2\pi c/\omega$. The black dashed line is the expected analytical scaling (Equation~\eqref{eq:lfac_equator_vs_r}). 
Right: The maximum Lorentz factor of the equatorial monster shock as a function $c \sigma_\times/\omega R_\times$. The wave amplitude $E_{\rm{w}}$ (green circles), wave frequency $\omega$ (blue diamonds), and magnetization $\sigma_\times$ (red squares) are varied. The black dashed line is a linear function with slope set by the analytical expectation (Equation~\eqref{eq:lfac_equator}) and a fitted non-zero $y$-intercept.  Details of the spherical 
FMS wave simulations can be found in Table~\ref{tab:sims}. All values are taken on the equator, $z/R_{\times}=0$. Individual linear fits where a single parameter is varied are presented in Appendix~\ref{app:fitting}, showing excellent agreement with the analytical expectation.}
    \label{fig:lfac_summary}
\end{figure*}

A single-fluid MHD description of the monster shock, as employed in this paper, holds generally true for $ 1/\Gamma_u \gg 2(\omega/\omega_\times)^{1/2}$ \citepalias{2023ApJ...959...34B}, where $\Gamma_u$ is the upstream Lorentz factor of the shock, $\omega$ is the FMS frequency, and $\omega_\times$ is the cyclotron frequency at the shock radius. That is, as long as the timescale of particle acceleration in the shock is much longer than the gyration timescale in the fluid frame, the assumptions of single-fluid MHD hold.
For a 10 kHz FMS wave with isotropic luminosity $L=10^{42} \,\rm{erg}\,\rm{s}^{-1}$ and a typical magnetar with magnetic moment $\mu=10^{33}\,$G$\,$cm$^3$, period $P=1\,$s, and multiplicity $\mathcal{M} = 10^6$, 
\begin{equation}
    1/\Gamma_u \sim 
    5\times10^{-5} \mathcal{M}_6 \mu_{33} L_{42}^{-1} \omega_5 P_0^{-1},
\end{equation}
and
\begin{equation}
    2(\omega/\omega_\times)^{1/2} \sim 
    2\times 10^{-5} (w_5 \mu_{33}^{1/2} L_{42}^{-3/4})^{1/2}.
\end{equation}
Beyond this limit, where the electron-positron pairs become unmagnetized, it has been argued that the wave
evolution remains similar to the 
single-fluid MHD description, albeit with modified jump conditions \citepalias{2023ApJ...959...34B}.

\begin{figure*}
    \centering
\includegraphics[width=1.0\textwidth]{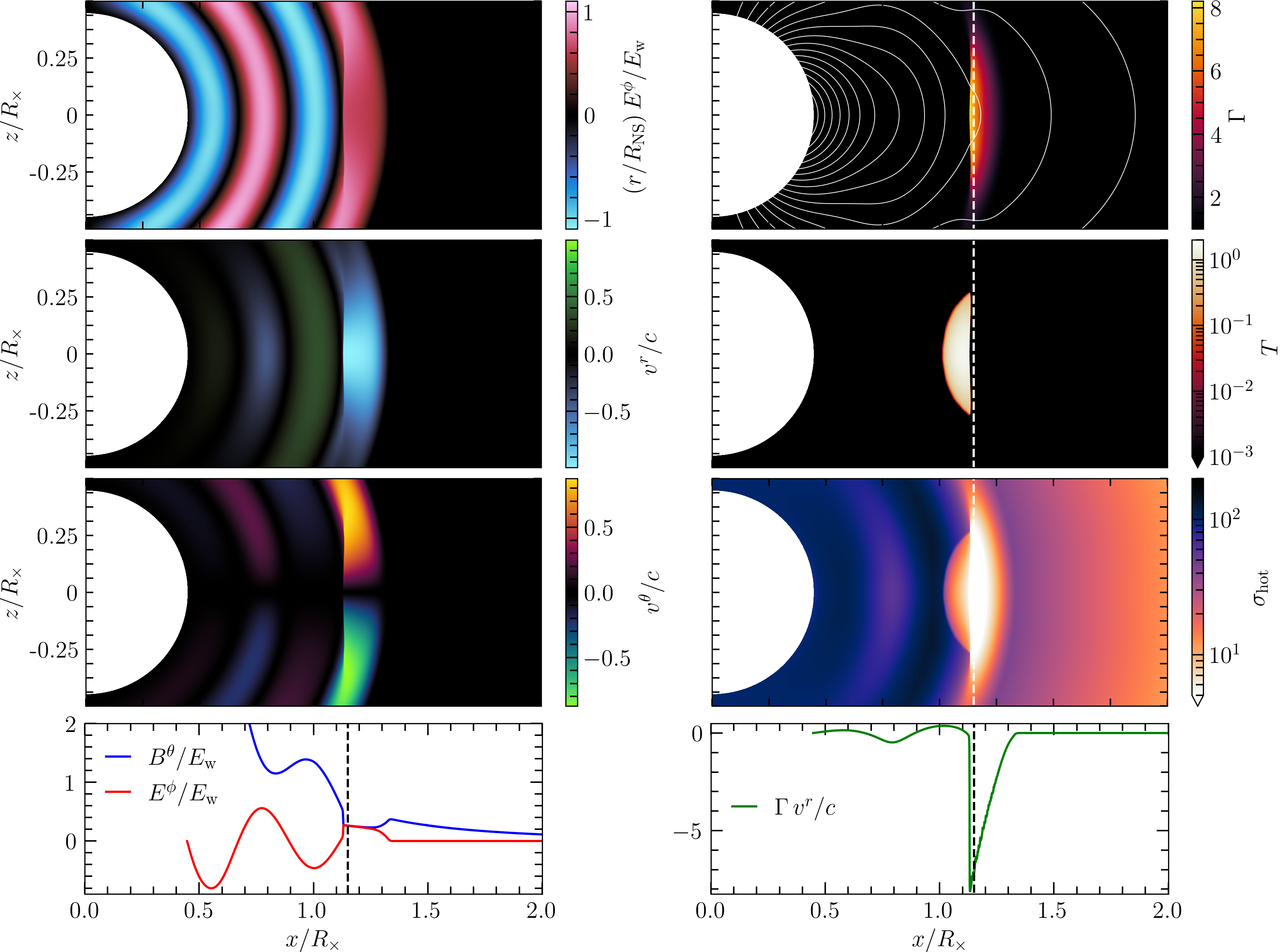}
    \caption{
    Visualizations of spherical FMS wave simulation \texttt{sigma100} at $t = 2  R_{\rm{NS}}/c$. 
    Top three rows are two-dimensional (2D) visualizations of: the toroidal ideal electric field $E^\phi$ (top left); the radial ${\bm E}\times{\bm B}$ velocity $v^r$ (middle left); the angular velocity $v^\theta$ (bottom left); the Lorentz factor $\Gamma$ (top right); the fluid temperature $T$ (middle right); and the magnetization accounting for the fluid temperature $\sigma_{\rm{hot}}$ (bottom right). Contours of poloidal magnetic flux are plotted as white solid lines; the contours are linearly spaced. A white (black) vertical dashed line is placed at $x=1.15R_\times$.
    Bottom row shows 1D equatorial slices 
    ($z/R_{\times}=0$) of: the FMS electromagnetic wave components $B^\theta$ and $E^\phi$ (left); and the radial four velocity (right). 
    }
    \label{fig:mshock}
\end{figure*}

\begin{figure}
    \centering
    \includegraphics[width=\linewidth]{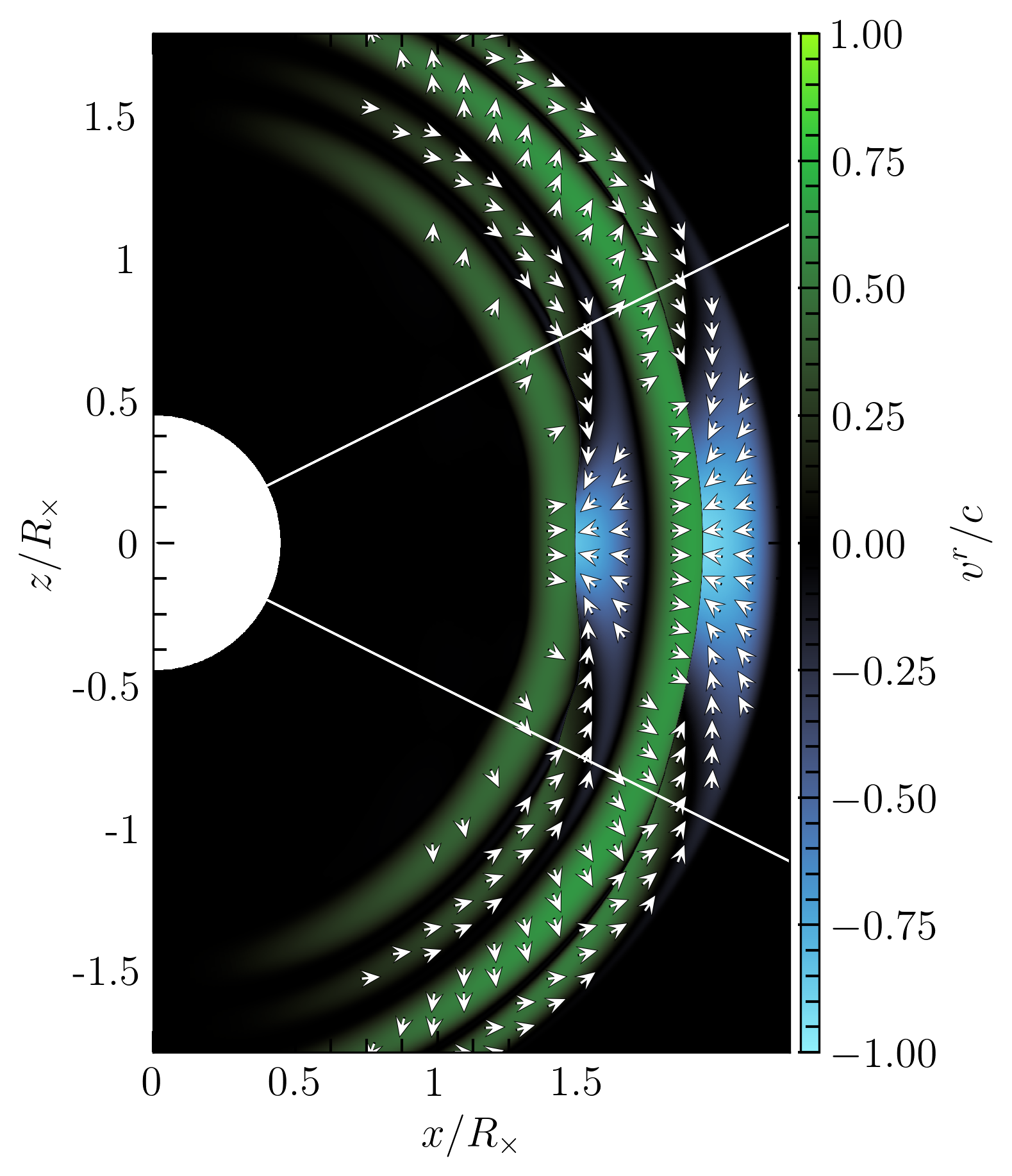}
    \cprotect\caption{The radial ${\bm E}\times{\bm B}$ velocity at $t=4 R_{\rm{NS}}/c$ in spherical FMS wave simulation \Verb+sigma100+. White arrows show the direction of flow in regions with Lorentz factor over a fixed threshold. Solid white lines are placed at $\sin\theta = \pm 2/\sqrt{5}$, the angle at which the radial velocity is expected to change sign for the  $\bm{B}_{\rm w} \cdot \bm{B}_{\rm bg} < 0$ portion of the wave.}
    \label{fig:vr}
\end{figure}

\begin{figure}
    \centering
    \includegraphics[width=\linewidth]{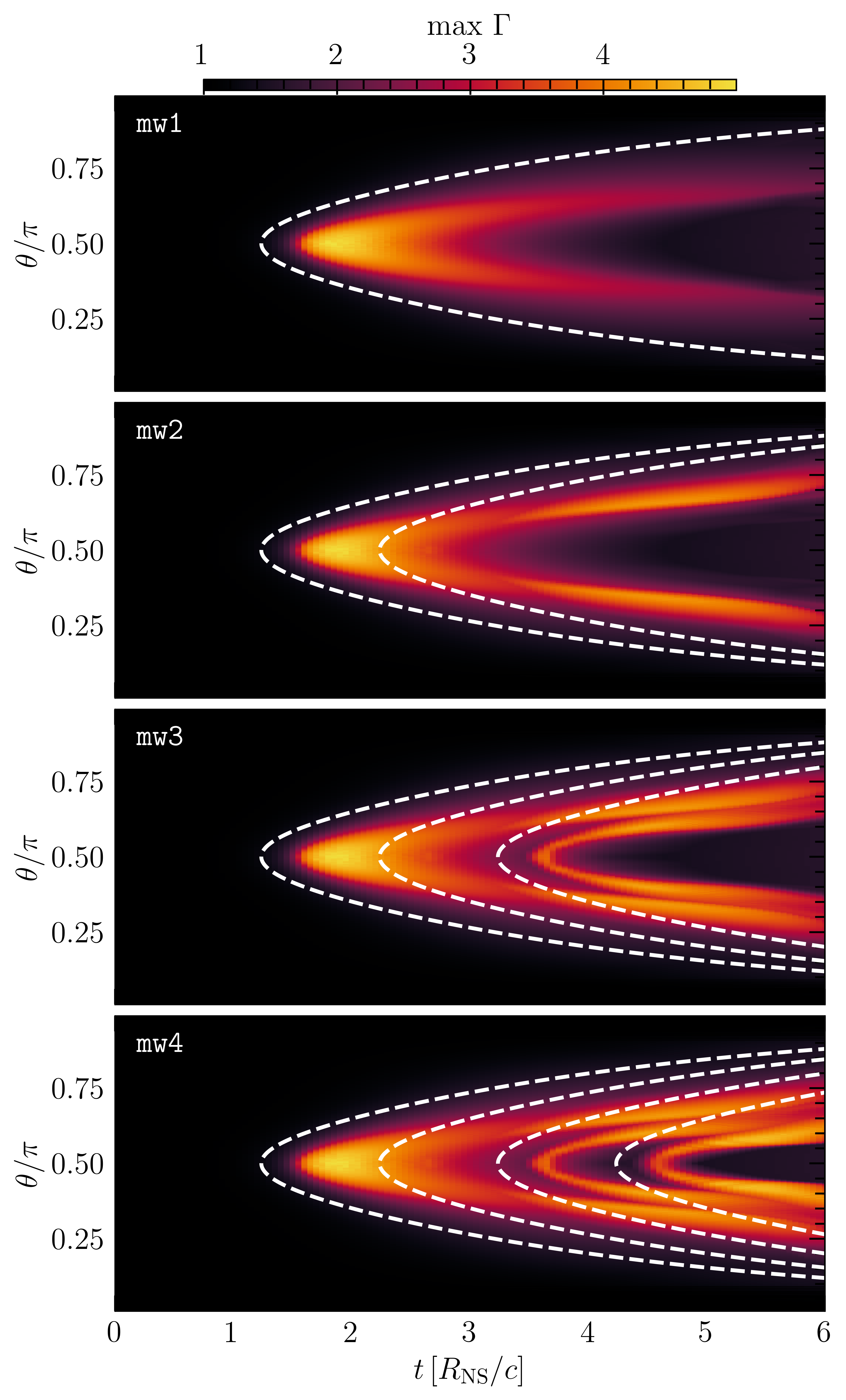}
    \cprotect\caption{The maximum ${\bm E}\times {\bm B}$ Lorentz factor, as a function of time and polar angle, for the spherical FMS wave simulations listed in Table~\ref{tab:sims};  these vary the number of FMS wavelengths launched from one to four.  The Lorentz factor in the secondary shock fronts is seen to peak off the equator.
    The analytical expectation for the non-linearity radius, $r_\times(\theta)$ (Equation~\eqref{eq:rx}),
    is plotted as a dashed white line for each wavelength, assuming propagation at the speed of light. 
    }
    \label{fig:lfac_2d_multi}
\end{figure}

\begin{figure*}
    \centering
    \includegraphics[width=\linewidth]{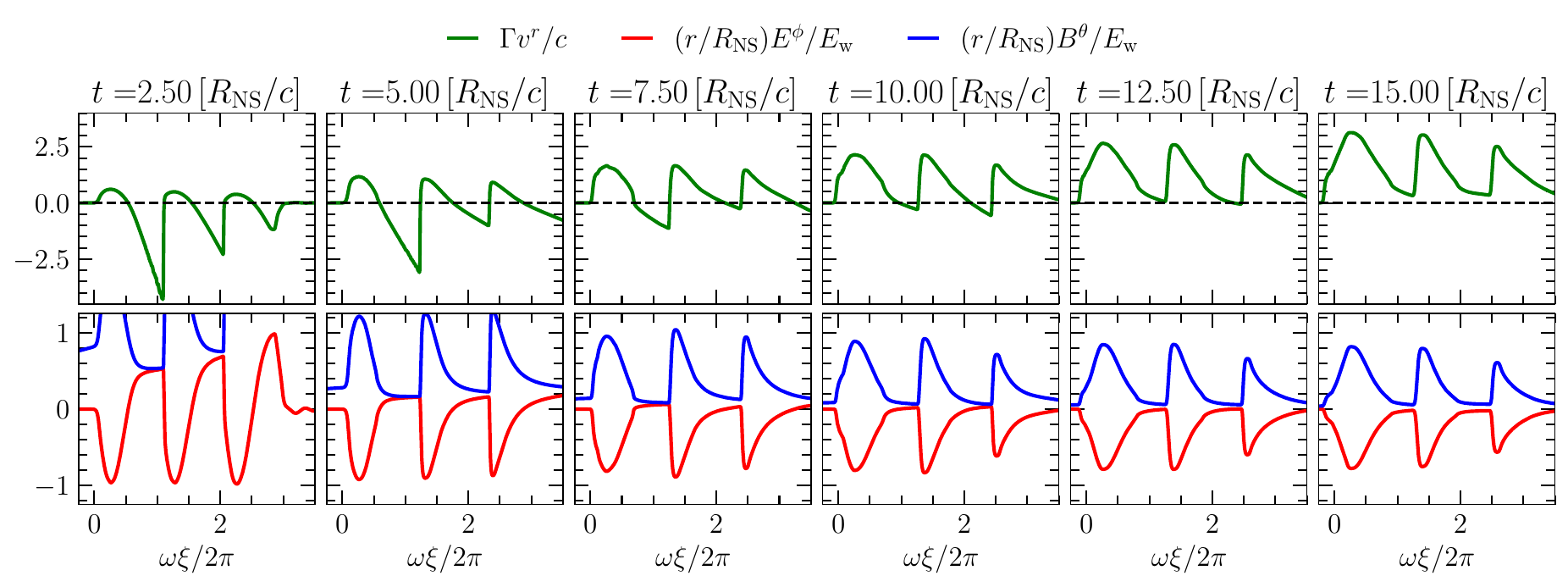}
    \caption{The on-equator time evolution of the multiple wavelengths launched in \texttt{mw3opp}. Top row: the radial four velocity, 
    with $v^r =0$ indicated by a dashed black line. Bottom row: the magnetic field component $B^\theta$ (blue) and electric field component $E^\phi$ (red). At late times a forward shock gradually forms at the front of the wavetrain, while the monster shock transitions from being inflow-dominated $(v^r <0)$ to outflow-dominated $(v^r>0)$. Panels are plotted as a function of the light cone coordinate $\xi = t - (r-R_{\rm NS})/c$ 
    (corresponding to stellar surface to the right and front of the wavetrain at $\xi=0$).
    }
    \label{fig:forwardshock}
\end{figure*}


\section{Quantitative Study of the Spherical 
Fast Magnetosonic Wave}\label{sec:results1}


Our MHD simulations of an expanding spherical FMS wave in a background dipole
(Table~\ref{tab:sims}) will now be compared with the quantitative prediction of monster 
shock formation in \citetalias{2023ApJ...959...34B}.  The overlap of the FMS wave and the background
dipole implies the formation of a surface in the magnetosphere where ${\bm E}\times{\bm B}$
drift becomes extremely relativistic.  This singularity is removed by inertial effects, which
are computable near the equator using the method of characteristics.  A key result is that in the short wave $(\lambda/R_\times\ll 1)$ and highly magnetized regime $(\sigma_\times \gg 1)$,
a FMS wave of frequency $\omega$ triggers a backflow toward the star with a Lorentz factor peaking
at \citepalias{2023ApJ...959...34B}
\begin{equation}
    \Gamma = \frac{\sigma_\times c}{\omega R_\times}. \label{eq:lfac_equator}
\end{equation}
This marks the formation of a monster shock near the nonlinearity radius $R_\times$.  Beyond this
point, the upstream Lorentz factor is expected to decay as
\begin{equation}
    \Gamma(x) = \frac{\sigma_{\rm bg} c}{\omega R_\times}  \left(\frac{R_\times}{x}\right) ^4\left[  \pi - 2 \arcsin \left(\frac{R_\times}{x}\right)^2 \right]. \label{eq:lfac_equator_vs_r}
\end{equation}

In the right panel of Figure~\ref{fig:lfac_summary}, the equatorial maximum upstream Lorentz factor is plotted as a function of magnetization, wave amplitude, and wave frequency. The simulation results reproduce the expected linear scaling of $\Gamma$ with $\sigma_{\times}$ and inverse scaling with frequency.
The left panels of Figure~\ref{fig:lfac_summary} show the upstream Lorentz factor
as a function of radius for three values each of magnetization and FMS frequency. The Lorentz factor
peaks slightly before the expected peak of $x/R_\times \sim 1.15$ and decays more slowly
beyond a radius $\sim 2R_\times$ than the analytical scaling (Equation~\eqref{eq:lfac_equator_vs_r}). 
Global PIC simulations of the monster shock have recently reported similar results \citep{2025arXiv250604175B}.
These include a modest  increase in the slope $d\ln\Gamma/d\ln\sigma_{\rm bg} \simeq 1.3$ 
on the equator.\footnote{The fit to $\Gamma(\sigma_{\times},\omega)$ in the PIC simulations was originally 
performed by varying only the slope and 
fixing the intercept to zero.  When the intercept is allowed to vary, a slope of near unity is obtained 
(Bernardi, private communication).} 

Note that the decay plateaus once the upstream Lorentz factor of the monster shock has become sufficiently close to unity.
It is important to note that the peak in the radially dependent piece of Equation~\eqref{eq:lfac_equator_vs_r} at $x/R_\times \sim 1.15$ is $\sim 0.82$ which is taken to be approximately unity to arrive at 
Equation~\eqref{eq:lfac_equator}. In the right panel of Figure~\ref{fig:lfac_summary} we find excellent agreement with a slope of unity as opposed to $\sim 0.82$ which would be too shallow. 
Some discrepancy is expected as Equation~\eqref{eq:lfac_equator_vs_r} was derived in the short wave, highly magnetized limit, while the simulations presented here are limited in both magnetization and wavelength. In Appendix~\ref{app:cyl}, we show that in an alternative and simplified quasi 1D geometry with $50-500$ times higher magnetization, that the discrepancy arises from pre-acceleration upstream of the plateau. This pre-acceleration arises because the plateau width is estimated under the assumption of force-free propagation and solving for $E = B = B_{\rm bg}/2$, thereby neglecting the upstream region where the Lorentz factor remains significantly below its peak value.
In the left panel of Figure~\ref{fig:lfac_summary}, the Lorentz factor is divided by its maximum value removing any sensitivity to this coefficient and instead isolating the radial dependence.

Visualizations of the simulation with highest magnetization, 
\texttt{sigma100}, are shown in Figure~\ref{fig:mshock}.   
The downstream of the shock 
is strongly heated as plasma accelerates inward relativistically (left panels).  
The half of the wave which forms the monster shock, with $\bm{B}_{\rm w} \cdot \bm{B}_{\rm bg} < 0$ and $E^\phi > 0$ (top and bottom left panels), is gradually shaved off;  eventually half the wave energy is converted into the kinetic and thermal energy of the plasma
(bottom left panel and right column).  The magnetization, defined in terms of the specific enthalpy $h$ and fluid frame magnetic field strength $b = B/\Gamma$, 
\begin{equation}
    \sigma_{\rm{hot}} = \frac{b^2}{4\pi \rho h c^2},
\end{equation}
is small both upstream, where the Lorentz factor is large, and downstream, where the enthalpy density is large. As expected, the plasma magnetization is smallest just upstream of the shock \citepalias{2023ApJ...959...34B}.

Off the equator, the shock becomes oblique and eventually disappears at a finite latitude. 
This effect is more visible in the expanded view of Figure~\ref{fig:vr}.
The flow pattern and the shape of the shock can be determined by overlapping the FMS wave with the magnetic dipole.
Singularity first develops at a radius \citepalias{2023ApJ...959...34B}
\begin{equation}
    r_\times(\theta) = R_\times \sqrt{\frac{4 - 3\sin^2\theta}{\sin\theta}}. \label{eq:rx}
\end{equation}
The drift velocity 
upstream of the shock, in the zone where ${\bm E}\times{\bm B}$ drift is still trans-relativistic, is
\begin{equation}\label{eq:br_theta}
    \bm{\beta}_{\rm{D}} = \frac{(4-5 \sin^2\theta) \hat{r} + 2\sin(2\theta) \hat{\theta}}{4-3\sin^2\theta}.
\end{equation}
Figure~\ref{fig:vr} shows the radial velocity and direction of the drift (white arrows) in 
simulation \texttt{sigma100}.  
An equatorial funnel is formed upstream of the shock, in which plasma drifts toward the equator, as well as being sucked toward the star. 
The radial velocity reverses sign at colatitude $\sin\theta = \pm 2/\sqrt{5}$ (as marked by diagonal lines);
one observes this pattern only in the zones upstream of the (two) shocks, as we expect from Equation~\eqref{eq:br_theta}.


We analyze the oblique portion of the shock while varying the number of wavelengths launched from the stellar boundary
(Figure~\ref{fig:lfac_2d_multi}).
Where the shock is no longer perpendicular to the ${\bm E}\times{\bm B}$ drift, the shock Lorentz factor is reduced.  
(In our simulations, this effect is caused partly by the decrease of magnetization with radius.)
The wavefront following the leading wave sees a perturbed background where plasma has been dragged and funneled toward the equator
by the leading wave. 
The subsequent wave sees a reduced magnetization on the equator and an enhanced magnetization off the equator, therefore forming two off-equator peaks in shock strength.  
Each wavefront leaves behind a perturbed magnetosphere by displacing plasma and changing the magnetization profile. The range of latitude 
over which the leading shock front will have developed by the time the second wave front steepens will depend directly on the wavelength 
of the wave train. Longer wavelengths allow for more angular spread, and thus a wider funnel of dense hot plasma, thereby pushing
the peak in the trailing shock 
further from the equator.

\subsection{Forward Shock}

A forward shock eventually forms at the head of a FMS wavetrain moving outward through a slowly rotating
magnetosphere.  When the leading part of the wave is compressive
($\bm{B}_{\rm w} \cdot \bm{B}_{\rm bg} > 0$), this shock develops at a radius \citepalias{2023ApJ...959...34B}
\begin{equation}\label{eq:shockrad}
R_{\rm F} = \left( \frac{8 c \sigma_\times}{\omega R_\times}\right)^{1/6} R_{\times};
\end{equation}
here, the outer magnetization profile (Equation~\eqref{eq:sigvsr}) has been assumed.
Well beyond this radius, a relativistic blast wave forms 
carrying $\sim 10\%$ of the initial wave energy \citep{2024ApJ...975..223B}.  

The evolution of a spherical FMS wavetrain with initial compression in a dipolar magnetic field is shown in Figure~\ref{fig:forwardshock}
(model \texttt{mw3opp}).  Here the magnetization is uniform in the outer magnetosphere, in which case Equation~\eqref{eq:shockrad} is
modified to
\begin{equation}
R_{\rm F} = \left( \frac{8 c \sigma_\times}{\omega R_\times}\right)^{1/3} R_{\times}.
\end{equation}
This works out to $R_{\rm F} = 5.4\,R_{\rm NS}$ for the simulation parameters $\sigma_{\rm bg} = 50$ and $\omega = 4\pi c/R_{\rm NS}$,
in reasonable agreement with the behavior shown in Figure~\ref{fig:forwardshock}.

Two strong and one weak on-equator monster shocks are formed in the simulation \texttt{mw3opp}.  Around the same radius where the forward shock forms,
one observes the bulk of the wave train transitioning to an explosive configuration.   In particular, 
the velocity jump across the internal monster shocks is observed to switch
from one dominated by inflow upstream of the shock, to a downstream outflow
(see the top panels in Figure~\ref{fig:forwardshock}).  A more detailed study of this effect would require substantially larger $\sigma_{\rm bg}$ and more advanced numerical schemes that remain stable at extreme magnetization.

\begin{figure*}
    \centering
    \includegraphics[width=1.0\textwidth]{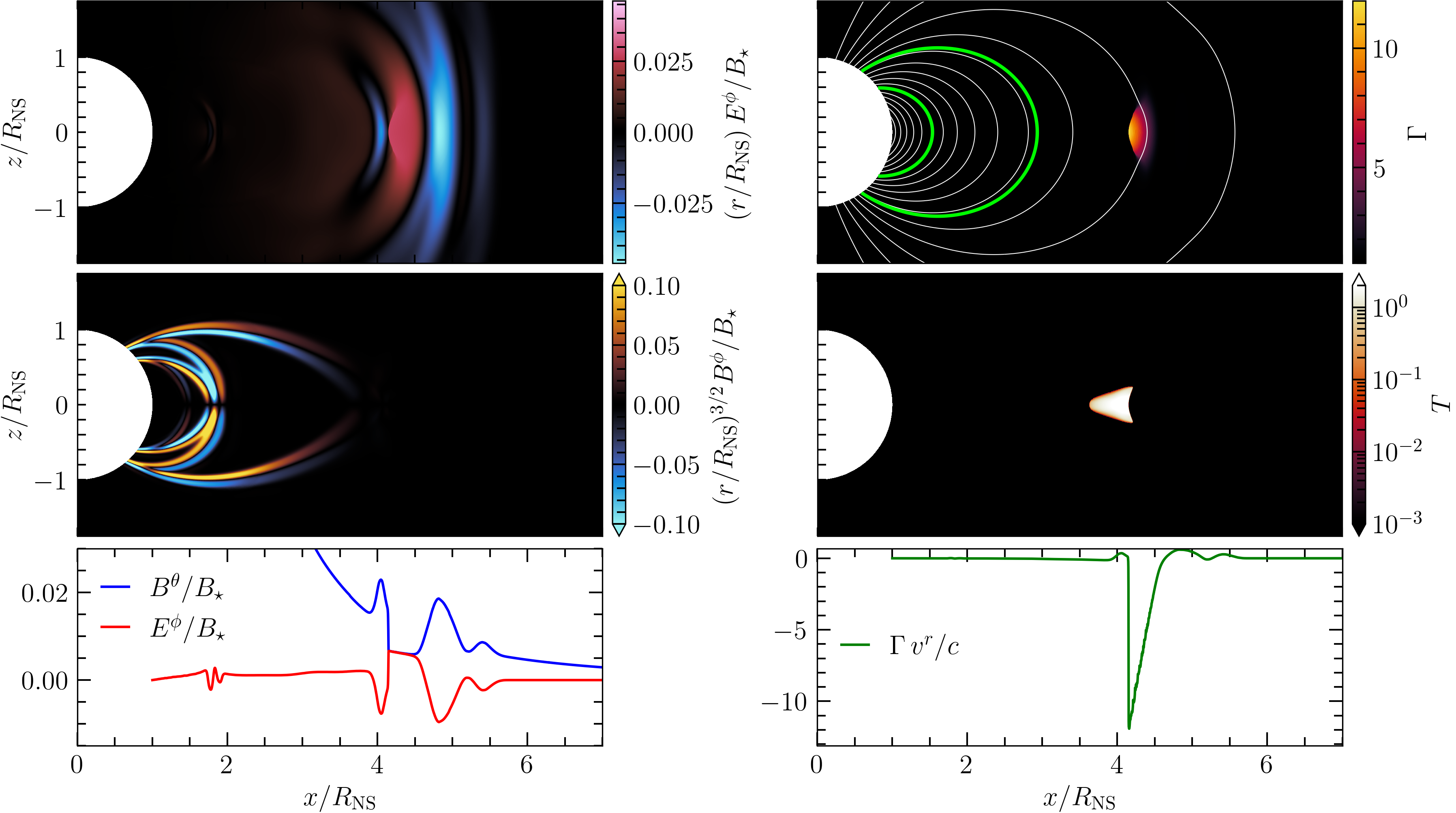}
    \caption{
    Visualizations of ${\rm A}\rightarrow{\rm FMS}$ mode conversion simulation \texttt{AW032pi} at $t = 5.0 \, [R_{\rm{NS}}/c]$. 
    Top two rows are 2D visualizations of: the toroidal ideal electric field $E^\phi$, indicating the FMS wave (top left); the toroidal magnetic field $B^\phi$, indicating the A waves (bottom left); the Lorentz factor $\Gamma$ (top right); and the fluid temperature $T$ (bottom right). Contours of poloidal magnetic flux are plotted as white solid lines; the contours are linearly spaced. 
    Green contours are drawn showing the magnetic field lines anchored at the edge of the twist.
    Bottom row shows 1D profiles of a monster shock forming on the equator ($z/R_{\rm NS}=0$):  the electromagnetic 
    field components $B^\theta$ and $E^\phi$ of the generated FMS wave (left); and the radial four velocity 
    (right). 
     }
    \label{fig:awshock}
\end{figure*}

\subsection{Cylindrical FMS}

In Appendix~\ref{app:cyl} we present quasi-1D isotropic FMS monster shock simulations in cylindrical geometry $(r,\phi,z)$, with background field $\bm{B}_{\rm bg} = B_\star (R_{\rm NS}/r)\hat{\phi}$ and wave field $\bm{E} = E_{\rm w}\sin(2\pi r/\lambda)\sqrt{R_{\rm NS}/r} \hat{z}$ \citep{2022arXiv221013506C}. We repeat the analysis of \citetalias{2023ApJ...959...34B} in this cylindrical setup, which may be viewed as a toy model for FMS waves steepening in the winds of rapidly rotating compact objects such as black holes, magnetars, or pulsars \citep{2025PhRvL.134c5201V}.
In this geometry the wave amplitude grows relative to the background field as $B_{\rm w}/B_{\rm bg} \propto \sqrt{r}$, leading to enhanced steepening at large radii. Owing to the purely perpendicular shock geometry and increased available resolution, we reach $\sigma_\times = 5\times 10^4$ with $\lambda/R_\times = 0.034$, exceeding the maximum magnetization achieved in kinetic simulations by over an order of magnitude (e.g., $\sigma_\times = 2560$ in \citealt{2025arXiv250604175B}) and surpassing previous fluid monster shock studies by multiple orders of magnitude \citep{2024ApJ...974L..12M}.
At these extreme magnetizations we find that the upstream Lorentz factor scales linearly with $c\sigma_\times/\omega R_\times$, where $R_\times$ is modified to account for the radial scalings of the background and wave fields. The measured slope of $\Gamma(r)$ agrees with the analytical prediction, while the plateau width exceeds the analytical prediction resulting in a modest pre-acceleration.

\section{The Perturbed Magnetosphere}\label{sec:pertresults}

We now turn to consider an indirect (but possibly more efficient) mechanism for sourcing a low frequency compressive wave
in a magnetar magnetosphere.  Alfv\'en waves are launched by twisting motions around the stellar poles.  They are a source
of FMS waves when they propagate along the curved magnetic field, or collide with oppositely propagating modes
\citep{2004PhRvD..70l4030T, 2021JPlPh..87f9014T, 2021ApJ...908..176Y, 2025ApJ...980..222B, 2025ApJ...995L..57B}.
The presence of high-frequency A waves 
(or secondary high-frequency FMS waves) creates zones of constructive interference with a FMS wave 
and therefore widens the zone of caustic formation.
The surface where $E^2-B^2 \rightarrow 0$ also can break up into independent patches, with more than one shock sometimes appearing in a given
direction.

\subsection{Alfv\'{e}n Mode Conversion}\label{sec:results2}

In the simulations summarized in Table~\ref{tab:sims_AWAW}, A waves are launched from a localized twist in the surface (Equation~\eqref{eq:twist}), which is mirrored in the opposing hemisphere with the same phase but opposing 
polarity.  The A waves meet at the equator, where they interact and launch a FMS wave.  This
non-isotropic compressive wave proceeds to steepen and produces a monster shock.  
A snapshot of this process is shown in Figure~\ref{fig:awshock}.  
A shock forms in the portion of the wave with $\bm{B}_{\rm w} \cdot \bm{B}_{\rm bg} < 0$,
driven by a 
huge inward-directed four velocity (see the lower two panels of Figure~\ref{fig:awshock}). 
At this phase of the expansion, heating is concentrated behind a cone with high Lorentz factor near
the equator.

Multiple simulations were performed with the amplitude and frequency of the localized twist varied to probe how the monster shock depends on the details of the dynamic twisting. Larger amplitude and lower frequency twisting results in stronger shocks, as expected from Equation~\eqref{eq:lfac_equator}. Exact scaling relationships between twist parameters and the upstream Lorentz factor require further investigation.

\subsection{Wrinkled Background}\label{sec:results3}

The presence of a spectrum of additional electromagnetic modes around a magnetar is a natural consequence of energy
release from the magnetic poles, e.g., the same process driving the conversion ${\rm A} + {\rm A} \rightarrow {\rm FMS}$.
Four simulations (Table~\ref{tab:sims_wiggles}) explore the effect of introducing a harmonic standing wave (a wrinkle)
into the initial conditions (Equation~\eqref{eq:wiggles}).

\begin{figure}
    \centering
    \includegraphics[width=\linewidth]{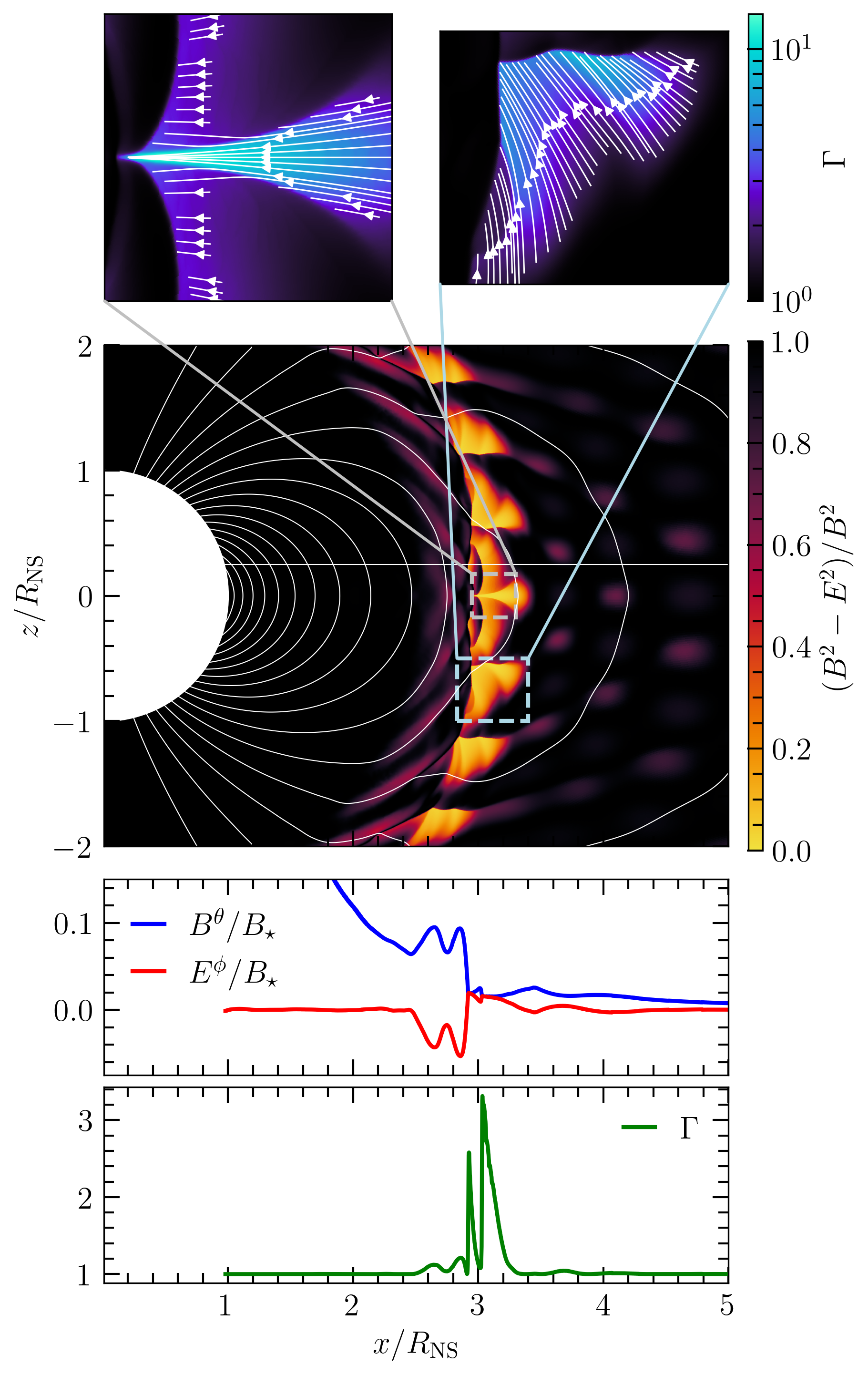}
\cprotect\caption{Visualizations of simulation \texttt{Wig0532} with perturbed background
at $t=2.5 R_{\rm{NS}}/c$, showing fragmentation of the shock front
due to the presence of large-amplitude standing waves in the magnetosphere.  
Top: 2D visualizations of the Lorentz factor $\Gamma$, zoomed in on regions where the shock has formed in the middle plot.  White lines show the flow stream lines upstream of the shock (in zones with sufficiently large four velocity).
Middle: 2D visualization of $(B^2 - E^2)/B^2$;  
regions approaching zero indicate the formation of a monster shock.  Solid white lines show contours of 
poloidal magnetic flux, with linear spacing.  Bottom two rows show 1D profiles of a double-shock structure forming at $z/R_{\rm NS}=0.25$: the electromagnetic field components $B^\theta$ and $E^\phi$; and the Lorentz factor $\Gamma$.
    }
    \label{fig:wrinkle_pretty}
\end{figure}

When the wrinkles are comparable in strength 
to the FMS wave, the shock is observed to fragment into disjoint pieces around the non-linearity radius $R_\times$;
see the middle panel of Figure~\ref{fig:wrinkle_pretty}.  The shock also appears closer to the star, in response to
the formation of zones of constructive interference, within which the peak Lorentz factor can increase by a factor $2-3$.

\begin{figure*}
  \centering
    \includegraphics[width=\linewidth]{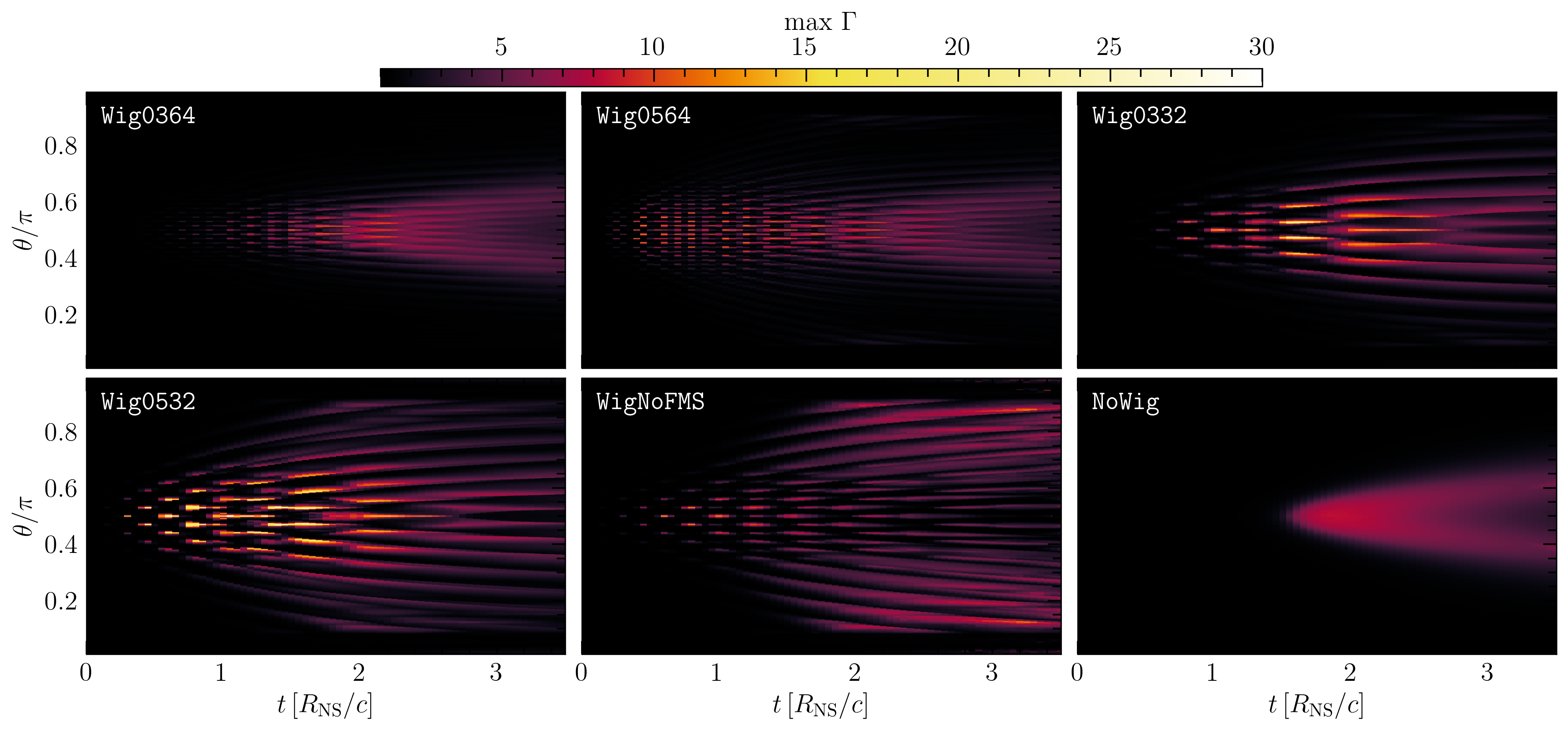}
  \caption{
  The maximum Lorentz factor along rays of constant polar angle $\theta$, as a function of time, for the 
  simulations listed in Table~\ref{tab:sims_wiggles}. Shock dissipation pushes inward to a smaller radius due to 
  constructive interference between the FMS wave and the large-amplitude wiggles.}
  \label{fig:wrinkle_lfac}
\end{figure*}

Another, more subtle, effect is the appearance of more than one shock along some lines of sight (bottom panel
of Figure~\ref{fig:wrinkle_pretty}).  
The appearance of a second shock induced by the wrinkle field $\{{\bm E}',{\bm B}'\}$
will lower the magnetization further back, to a value
$\sigma_{\rm eff} \sim (B_d/B')^2$, where $B_d$ is the background dipole field. 

To understand this effect, we decompose the relativistic invariant ${\cal F} =
E^2 - B^2$ into contributions from the background dipole ${\bm B}_d$, the fast wave $\{{\bm E}_{\rm F},{\bm B}_{\rm F}\}$, and the wrinkle wave.
The wrinkle wave is most generally a superposition of A and FMS waves (both of which are seen 
in large numbers in the simulation of \citealt{2025ApJ...995L..57B}).  Because the fast and wrinkle fields are dynamic, dissipation is triggered not only by a
partial cancellation of the magnetic field, but also through
an overlap of the wrinkle electric field ${\bm E}'$ with the FMS field ${\bm E}_{\rm F}$.

Near the magnetic equator, the FMS wrinkle wave is partly compressive, 
${\bm B}'_{\rm F} = B'_r \hat r + B_\theta' \hat\theta$;  a non-axisymmetric A wave, if present, has magnetic field ${\bm B}'_{\rm A} = B_r' \hat r + B'_\phi\hat\phi$.
The invariant may be written, for small wrinkle amplitude, as
\begin{equation}
\begin{split}
{\cal F} & =  ({\bm E}_{\rm F} + {\bm E}')^2 - ({\bm B}_d + {\bm B}_{\rm F} + {\bm B}')^2 \\
&\simeq E_{\rm F}^2 - (B_{\rm F} + B_d)^2 + 2E_{\rm F} E'_\phi\\ & \qquad - 2B'_\theta (B_d + B_{\rm F}).
\end{split}
\end{equation}
Here, we have used the shorthand ${\bm B}_d = B_d\hat\theta$ and ${\bm B}_{\rm F} = B_F\hat\theta$ near the equator.

We are interested in an initial fast wave pulse with $\bm{B}_{\rm F} \cdot \bm{ B_d} < 0$.  We consider the state of the electromagnetic field
at some radius $r> R_\times$, where a part of the pulse
has plateaued with ${\cal F} \simeq 0$ \citepalias{2023ApJ...959...34B}; see also Figure~\ref{fig:forwardshock}.  
Near the outer edge of this plateau, which sits at light cone coordinate $\xi = t-r/c$ (with $\xi = 0$ denoting
the front of the wave), we may Taylor expand the fast wave fields as $E_{\rm F} = E_0 \omega_{\rm F} \xi$, 
$B_{\rm F} = -E_0\omega_{\rm F}\xi \simeq -{1\over 2}B_d$.  It will be simplest to consider a wrinkle field with vanishing 
$B'_\theta$ and radial wavenumber $k'_r \gtrsim \omega'/c$.  Then, at the plateau boundary, the derivative of ${\cal F}$ 
with respect to radius simplifies to
\begin{equation}\label{eq:invar}
\begin{split}
{d{\cal F}\over dr} & \simeq {d\over dr}\left(2E_{\rm F}E'_\phi - 2B_{\rm F}B_d\right)\\ & \simeq B_d(k'_rE'_\phi - 2\omega_{\rm F} E_0/c).
\end{split}
\end{equation}
Here we have assumed that $|k_r'| \gg \omega_F/c \gg 1/r$.  
The second term on the right-hand side of Equation~\eqref{eq:invar}
represents the transition from $E^2-B^2 \simeq 0$ to $E^2-B^2 < 0$ outside the FMS pulse.
The first term, due to the wrinkle,
allows ${\cal F}$ to develop a secondary maximum in some finite layer of thickness $\sim 1/k'_r$ 
around the outer edge of the plateau.
In this way, a second shock can form even if the wrinkle amplitude is small compared with the fast wave amplitude $E_0$,
as long as $k'_r E' > 2\omega_{\rm F} E_0/c$.  

The partial dissipation of a wrinkle field $B'$ through a second shock generates enthalpy of 
density $w' \sim {B'}^2/4\pi$ (where the fast wave and wrinkle interfere constructively with 
${\bm E}_{\rm F}\cdot{\bm E}' > 0$).  The magnetization drops as a result to $\sigma_{\rm eff} \sim (B_d/B')^2$;  in this way, the wrinkle field provides 
an effective inertia density ${B'}^2/4\pi c^2$ for the larger-scale MHD flow.



In the four simulations, we varied both the amplitude and the frequency of the wrinkles.  The angular and temporal dependence of the maximum Lorentz factor is shown in Figure~\ref{fig:wrinkle_lfac}. Larger amplitude, lower frequency standing waves resulted in a stronger enhancement of the Lorentz factor. Notably, \texttt{Wig0332} and \texttt{Wig0532} see the largest enhancement, indicating a strong dependence on the frequency of the standing modes. A run with no FMS, \texttt{WigNoFMS}, was also performed;  here, some localized peaks in the Lorentz factor still form but are much weaker than when combined with the FMS in runs \texttt{Wig0532} and \texttt{Wig0332}.

\begin{figure*}
    \centering
\includegraphics[width=\linewidth]{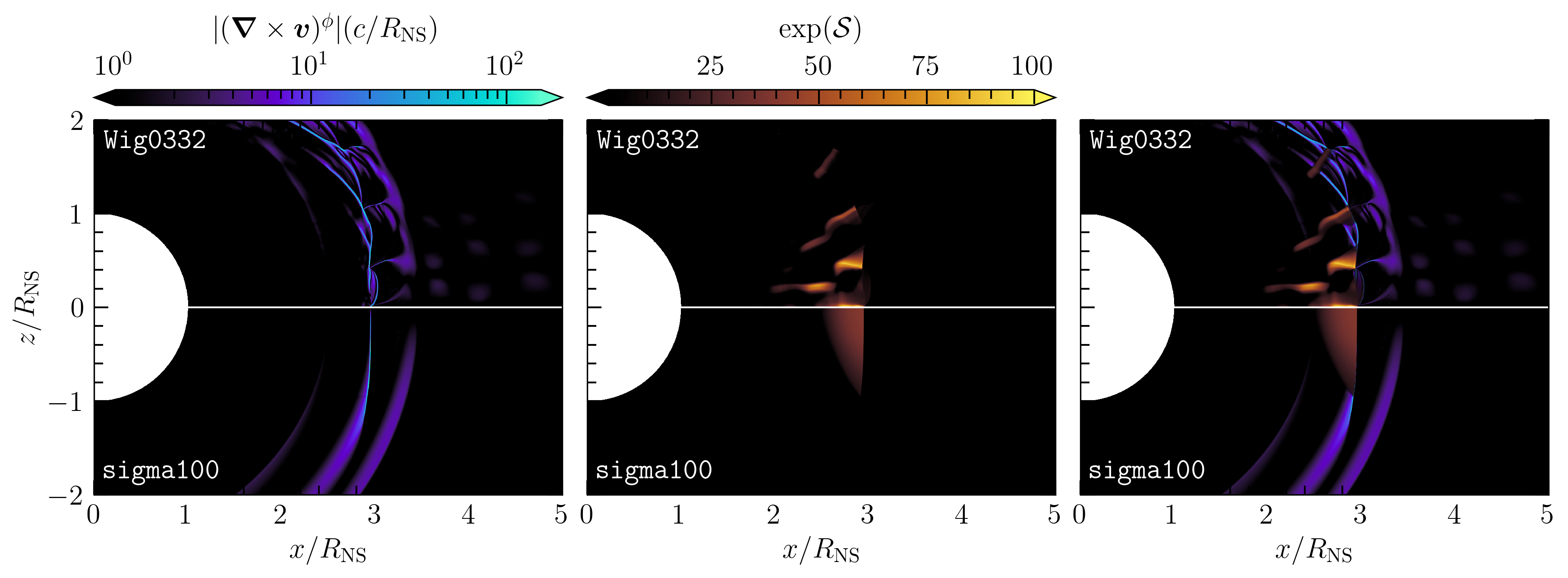}
    \caption{
    The vorticity (left), dimensionless entropy $\mathcal{S} = (s-s_0)/c_V=\ln(p/\rho^{-\hat{\gamma}})$ (middle) and both quantities overlaid (right) at $t= 2.5 R_{\rm{NS}}/c$
    in runs \texttt{Wig0532} and \texttt{sigma100}.  
    The simulation with perturbed background develops strong shear at the boundary of the fragmented shock patches 
    and along the shocks themselves.  In simulation \texttt{sigma100} with uniform background dipole, the vorticity peaks only along the shock. Entropy is generated at the shock in both cases, and appears around the boundaries of the shock patches in simulation \texttt{Wig0532} as a result of the shifting position of the shock.
}  \label{fig:vorticity+entropy}
\end{figure*}

The edges of fragmented shocks are sites of strong vorticity and entropy generation:  Figure~\ref{fig:vorticity+entropy} compares
simulation \texttt{Wig0532} with its unperturbed counterpart.  The narrow and intense vortex layers may be sites of secondary instabilies 
and particle acceleration;  further understanding would be aided by three-dimensional (3D) or kinetic simulations.

\section{Conclusion}\label{sec:conclusion}

We have presented 2D, axisymmetric, relativistic ideal MHD simulations of a strongly perturbed magnetic dipole field.  
The transformation of a breaking compressive wave into a monster shock has been demonstrated
for three families of initial conditions: spherical FMS waves, A mode collisions that convert to FMS waves 
around the magnetic equator, and FMS waves propagating through a wrinkled dipole.  

The novel MHD method developed for this investigation 
has allowed us to reach the asymptotic monster shock regime for the first time in a global MHD simulation.
The dissipation of electromagnetic energy into kinetic and thermal energy at high Lorentz factor has been described for
magnetizations as high as 100 in dipole geometry and $5\times 10^4$ in an axisymmetric, cylindrical test problem.
Wave breaking in an inhomogeneous background at high Lorentz factors leads to shock fragmentation and the formation
of secondary shocks and narrow vortex layers.  Extensive numerical tests are described in the Appendices.


Our simulations of a spherical FMS wave (Section~\ref{sec:results1}) successfully tested some key predictions of the 
analytic solution found by \citetalias{2023ApJ...959...34B}.  These include both the drift velocity profile established in the quasi-linear
regime and the Lorentz factor profile near the peak of shock dissipation.  We confirm the scaling of peak 
$\Gamma$ with background magnetization and FMS frequency, albeit with a small constant offset (Figure~\ref{fig:lfac_summary}).
We also find that the upstream Lorentz factor decays more slowly than predicted,
in agreement with results from recent kinetic simulations \citep{2025arXiv250604175B}. 
The deviations from the analytical predictions (derived in the limit of high $\Gamma$ and short wavelength $\lambda\ll r$) may be caused by the limited range of $\Gamma$ achieved in the simulations. The limitation becomes most significant at radii at $r\gg R_\times$ where $\Gamma$ decreases. 

When multiple FMS wavefonts are present, the peak of dissipation is found to
shift away from the equator for the secondary shocks (Figure~\ref{fig:lfac_2d_multi}), due to the decrease
in equatorial magnetization behind the primary shock.  We also observe a transition from inflow-dominated shocks
to an explosive configuration in which the star-frame Lorentz factor is highest behind each shock
(Figure~\ref{fig:forwardshock}).   When the initial part of the FMS pulse is compressive,
a forward shock is seen to develop, consistent with expectations \citepalias{2023ApJ...959...34B}.


Such waves may be directly injected into the magnetosphere via  magneto-elastic  quakes in the crust, albeit with low efficiency $\lesssim 4\%$ \citep{2026ApJ...998..190Q}, or with larger relative amplitude during collapse of a hypermassive neutron star \citep{2012PhRvD..86j4035L,2024ApJ...974L..12M}. 

A complementary set of simulations (Section~\ref{sec:results2}) probed the emission of a FMS wave in a magnetosphere 
experiencing quasi-periodic
torsional oscillations (see also \citealt{2024ApJ...972..139M, 2025ApJ...980..222B}).  The FMS wave is excited near the
equator by overlapping A modes and then is seen to propagate outward and steepen into a quasi-conical shock 
(Figure~\ref{fig:awshock}).  The geometry of the shock
so formed will depend on where the A modes collide and their relative amplitude, phase, and frequency.


The strongest modifications to the structure of the monster shock are observed in a magnetosphere where higher-frequency MHD modes
are already present in the wave breaking zone (Section~\ref{sec:results3}). 
These wrinkle modes interact both constructively and destructively with the FMS wave, producing a few related effects.
The shock fragments when the wrinkles have comparable amplitudes to the FMS wave 
and their wavenumbers are moderate (Figure~\ref{fig:wrinkle_pretty}). 
We also observe localized enhancements of the upstream Lorentz factor (Figure~\ref{fig:wrinkle_lfac}) and significant vorticity 
at the edges of the fragmented shock patches (Figure~\ref{fig:vorticity+entropy}). 
In the perturbed configuration, entropy is generated both in the shocks and at the interfaces between fragmented regions. 

Along any given line of sight, the shock can intermittently appear and disappear (Figure~\ref{fig:wrinkle_lfac}), and a second shock may form.
Then the magnetization experienced by the innermost shock is strongly reduced to $\sim (B_d/B')^2$ in the presence of a wrinkle
of amplitude $B'$.   The formation of secondary maxima in the relativistic invariant $E^2-B^2$ is possible when the 
radial wavenumber of the wrinkle is large enough that $k_r' B' \gg 2 \omega_F E_0/c$. A diverging current density spectrum
$k_\perp B'$ is a common feature of turbulent MHD cascades, and the formation of multiple shocks deserves further investigation.
The efficiency of X-ray emission from the shock may not be significantly changed, but
there are also interesting implications for the light curve of electromagnetic radiation produced by a monster shock.


 Some further insight into the effects of pre-existing magnetic wrinkles,
such as the appearance of secondary shocks, can be obtained with more focused 2D and 3D simulations.
Future 3D simulations will be needed to investigate the effect of non-linear mode interactions such as ${\rm A} + {\rm A} \leftrightarrow
{\rm FMS}$ on shock formation \citep{golbraikh2023}. 
Additionally, the relativistic ideal MHD model employed in this paper neglects additional physics, such as cooling and pair production, which are crucial for a full description of the monster shock, and will be added in future studies.  
Precursor emission, a kinetic effect, cannot be captured by MHD simulations.

\section*{Acknowledgments}

The authors thank James Beattie, Dominic Bernardi, Koushik Chatterjee, Daniel Gro\v{s}elj, Amir Levinson, Jens Mahlmann, Alexander
Philippov, Oliver Porth, Arno Vanthieghem, and Jonathan Zhang for insightful discussions. 

This research was supported in part by grant NSF PHY-2309135 to the Kavli Institute for Theoretical Physics (KITP).

M.~.P.~G. acknowledges the support of the Natural Sciences and Engineering Research Council of Canada (NSERC), [CGS D - 588952 - 2024]. Cette recherche a \'{e}t\'{e} financ\'{e}e par le Conseil de recherches en sciences naturelles et en g\'{e}nie du Canada (CRSNG), [CGS D - 588952 - 2024].

This research was supported by grant 23JWGO2A01 from the Canadian Space Agency (B.~R.), by grants 613413 and RGPIN-2023-04844 from NSERC (M.~P.~G. and B.~R.), by grant MP-SCMPS-00001470
from the Simons Foundation (B.~R. and C.~T.), and by NSERC grant RGPIN-2023-04612 (C.~T.).
B.~R. acknowledges a guest researcher position at the Flatiron Institute, supported by the Simons Foundation.  
A.M.B. acknowledges support
by NASA grant 80NSSC24K1229, NSF grant AST-2408199, and Simons Foundation grant 446228. This work was also facilitated by Multimessenger Plasma
Physics Center(MPPC) grant PHY-2206609.

E.~R.~M. acknowledges support by the National Science Foundation under grants No.
PHY-2309210 and AST-2307394, and from NASA's ATP program under grant
80NSSC24K1229.

The computational resources and services used in this work were partially provided by facilities supported by the VSC (Flemish Supercomputer Center), funded by the Research Foundation Flanders (FWO) and the Flemish Government – department EWI, by the Scientific Computing Core at the Flatiron Institute, a division of the
Simons Foundation, and by Compute Ontario and the Digital Research Alliance of Canada (alliancecan.ca) compute allocation rrg-ripperda.


\appendix

\section{Numerical Methods}\label{app:num_methods}

All simulations presented in this paper are performed using the Black Hole Accretion Code (\Verb+BHAC+) \citep{Porth:2016rfi, Olivares2019} which solves the multidimensional general-relativistic ideal magnetohydrodynamic equations. We perform exclusively ideal MHD simulations using the constrained transport method to evolve the magnetic flux and preserve $\bm{\del} \cdot \mbf{B}=0$ to machine precision \citep{Olivares2019}, where $B$ is the magnetic field. 
In this Section~we write the model in Heaviside–Lorentz units with $c=k_B=1$, $B/\sqrt{4\pi} \rightarrow B$, and $E/\sqrt{4\pi} \rightarrow E$, where $c$ is the speed of light, $k_B$ is the Boltzmann constant, and $E$ is the electric field. Additionally, we use the contravariant and covariant forms of vectors and tensors, as opposed to the physical components used throughout the main text. Note we use Latin indices to indicate spatial indices (e.g. $i=1,2,3$) and Greek indices to denote spacetime indices (e.g. $\mu=0,1,2,3$). For completeness, we state the model in the 3+1 Arnowitt–Deser–Misner (ADM) formalism \citep{1962gicr.book..227A}, with metric tensor
\begin{equation}\label{eq:metric_tensor}
    g^{\mu \nu} = \left(  
    \begin{array}{cc}
        -1/\alpha^2 & \beta^j/\alpha^2 \\
        \beta^j/\alpha^2 & \gamma^{ij} - \beta^i \beta^j/\alpha^2 
    \end{array}
    \right),
\end{equation}
where $\alpha$ is the lapse function, $\beta^i$ is the shift vector, and $\gamma^{ij}$ is the emergent spatial metric on the spacelike hypersurfaces. The simulations presented in this paper are in flat Minkowski spacetime with spherical coordinates $(r, \theta, \phi)$ with $\alpha =1$, $\beta^i = 0$ and 
\begin{equation}
    \gamma_{ij} = \left( 
\begin{array}{ccc}
    1 & 0 & 0  \\
    0 & r^2 & 0  \\
    0 & 0 & r^2\sin^2\theta
\end{array}
    \right).
\end{equation}
The ideal MHD equations can written in conservative form,
\begin{equation}
    \partial_t \left( \sqrt{\gamma} \bm{U} \right) + \partial_i \left( \sqrt{\gamma} \bm{F}^i \right) = \sqrt{\gamma} \bm{S},
\end{equation}
with conserved variables $\bm{U}$, conserved fluxes $\bm{F}$, and non-conserved fluxes including source terms, $\bm{S}$.
The conserved variables, fluxes and sources are defined as
\begin{align}
    &\bm{U}  = \left[ 
    \begin{array}{c}
        D   \\
        S_j  \\
        \tau \\
        B^j  \\
    \end{array}
    \right],  
    & \bm{F}^i & = \left[ 
    \begin{array}{c}
        \mathcal{V}^i D\\
        \alpha W^i_j - \beta^i S_j\\
        \alpha (S^i - v^i D) - \beta^i \tau\\
        \mathcal{V}^i B^j - B^i \mathcal{V}^j
    \end{array}
    \right],
    & \bm{S} = \left[ 
    \begin{array}{c}
        0\\
        \frac{1}{2} \alpha W^{ik} \partial_j \gamma_{ik} + S_i \partial_j \beta^i - U \partial_j \alpha\\
        \frac{1}{2} W^{ik} \beta^j \partial_j \gamma_{ik} + W^j_i \partial_j \beta^i - S^j \partial_j \alpha\\
        0\\
    \end{array}
    \right],
\end{align}
where $\mathcal{V}^i = \alpha v^i - \beta^i$ is the the transport velocity, $v^i$ is the spatial three-velocity, $p$ is the gas pressure, $D = \Gamma \rho$ is the lab frame mass density, $\Gamma$ is the Lorentz factor, $\rho$ is the mass density, and $\tau$ is the the (rescaled) conserved energy density $\tau = U - D$, with energy density
\begin{equation}
    U = \rho h \Gamma^2 - p + \frac{1}{2} \left( B^2(1 + v^2) - (B^j v_j)^2 \right),
\end{equation}
and the specific enthalpy is
\begin{equation}
    h = 1 + \frac{\hat{\gamma}}{\hat{\gamma}-1} \frac{p}{\rho},
\end{equation}
where $\hat{\gamma}$ is the adiabatic index.
The covariant momentum density is
\begin{equation}
    S_i = \rho h \Gamma^2 v_i + B^2 v_i - (B^j v_j) B_i,
\end{equation}
and the spatial variant of the stress-energy tensor is
\begin{equation}
\begin{split}
    W^{ij} = S^i v^j + (p +b^2/2) \gamma^{ij} - \frac{B^i B^j}{\Gamma^2} - (B^k v_k) v^i B^j,
\end{split}
\end{equation}
where $b^2 = B^2 - E^2$ is the
square of the fluid frame magnetic field strength, and $E^i = - \gamma^{1/2} \epsilon^{ijk} v_j B_k$ is the ideal electric field. Recovery of the primitive variables, upon which the fluxes and source terms depend, is performed using the 2D scheme with a 1DW fallback, as first described in \cite{2006ApJ...641..626N}, with our particular implementation outlined in \cite{Porth:2016rfi}.

An entropy switch \citep{2009ApJ...692..411N, 2013MNRAS.429.3533S, Porth:2016rfi} is employed as a backup when the primitive variable recovery fails and is automatically applied for regions below a threshold of $\beta = 2 p/B^2 \leq 10^{-2}$. Therefore, an entropy evolution equation is evolved alongside the above ideal MHD equations
\begin{equation}
    \partial_t (\sqrt{\gamma} \Gamma S)+\partial_i [(\sqrt{\gamma} \mathcal{V}^i) \Gamma S]=0,
\end{equation}
where $S = p/\rho^{\hat{\gamma}-1}$, $\Gamma S$ is the conserved quantity, and the entropy $s=p/\rho^{\hat{\gamma}}$ is the primitive quantity. This allows for the recovery of primitive variables without the use of the rescaled energy, $\tau$, which is typically the source of numerical failures in large $\sigma$ or low $\beta$ regions. When entropy evolution is employed, after the primitive variables are recovered the value of  $\tau$ is disposed of and a new value consistent with entropy evolution is evaluated. When energy evolution is employed, the entropy is updated following the successful recovery of primitive variable as to be consistent with energy conservation. For the simulations presented in this paper, due to the zero temperature background, entropy evolution is employed outside of both the shock and heated regions where the plasma is rapidly heated above the plasma $\beta$ cutoff. Thus, energy evolution is used in the shock, where entropy is expected to be generated. Entropy evolution outside of the shock and heated regions does not strictly conserve energy, this alongside the flooring and dissipation described in Appendix~\ref{app:amr_flooring} results in  $\sim0.4\%$ of the total energy being lost in the runtime of isotropic FMS wave simulation \texttt{sigma100}.
See \cite{Porth:2016rfi, Olivares2019} for further details on the ideal MHD model and numerical scheme implemented in \texttt{BHAC}.

\section{Adaptive Mesh Refinement, Flooring, and Numerical Stability}\label{app:amr_flooring}

\begin{figure*}
    \centering
    \includegraphics[width=\textwidth]{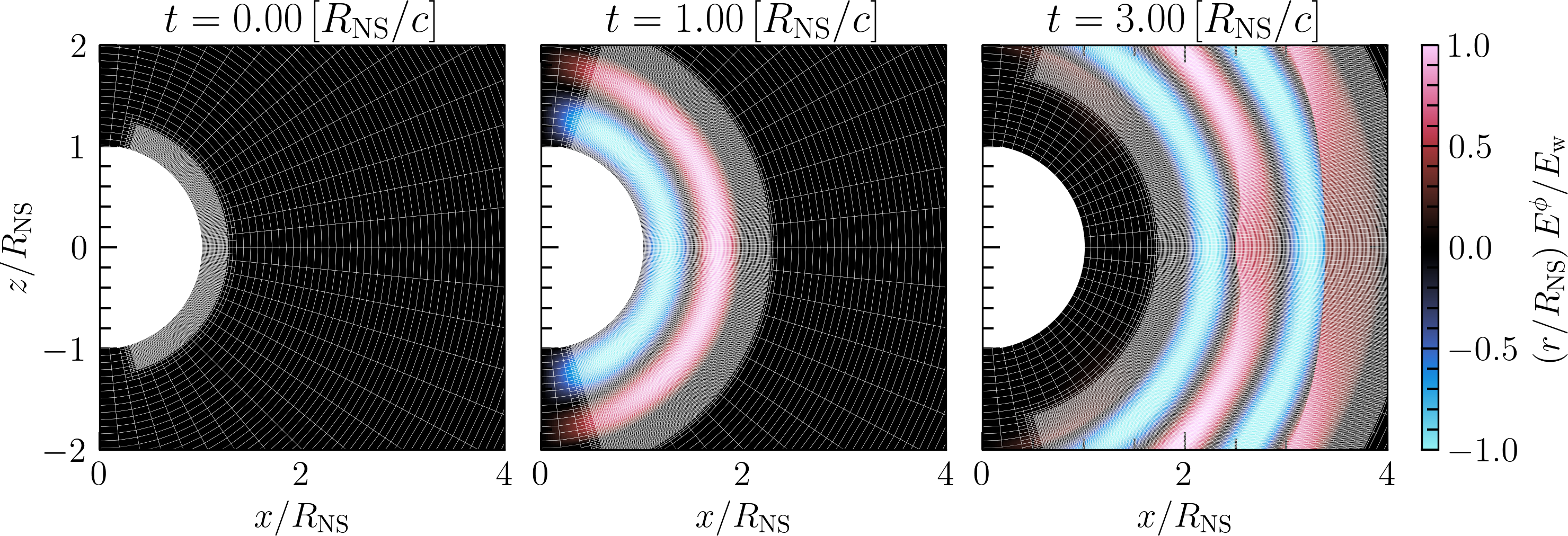}
    \caption{
    The toroidal component of the ideal electric field $E^\phi$, for an MHD monster shock initiated through the launching of an isotropic fast wave from the stellar surface with $\sigma_\times=100$. The outline of mesh blocks is shown in white, with each block containing 16 radial cells and 8 angular cells. The plotted simulation has domain $r/R_{\rm{NS}}\in[1,10]$ and $\theta\in[0,\pi]$ with a base grid of $2048\times256$ cells and 3 levels of mesh refinement for a maximum effective resolution of $16384\times2048$.} 
    \label{fig:amr}
\end{figure*}

To maximize resolution across the wave while minimizing computational expense we employ an AMR scheme which focuses resolution on the dynamically important portions of the domain. The grid utilized in all simulations presented in this paper is linear in the radial coordinate; this ensures that any numerical diffusion across the wave is constant in time.
For the isotropic fast modes this tracks the launched wavelengths in space. To do this we force the grid to the maximum AMR level for $ r_ - < r < r_+$, where
\begin{align}
    r_+ & = R_{\rm{NS}} + ct + 0.25 R_{\rm{NS}},\\
    r_- & = R_{\rm{NS}} + ct - N \lambda - 0.25 R_{\rm{NS}},
\end{align}
where $\lambda$ is the wavelength of the launched wave, and $N$ is the number of wavelengths launched. This tracking of the waves is shown in Figure~\ref{fig:amr} for $N=2$. Note that the mesh refinement does not cover the highest latitude portions of the wave to avoid the numerical pole. 

\begin{figure*}
    \centering
    \includegraphics[width=\textwidth]{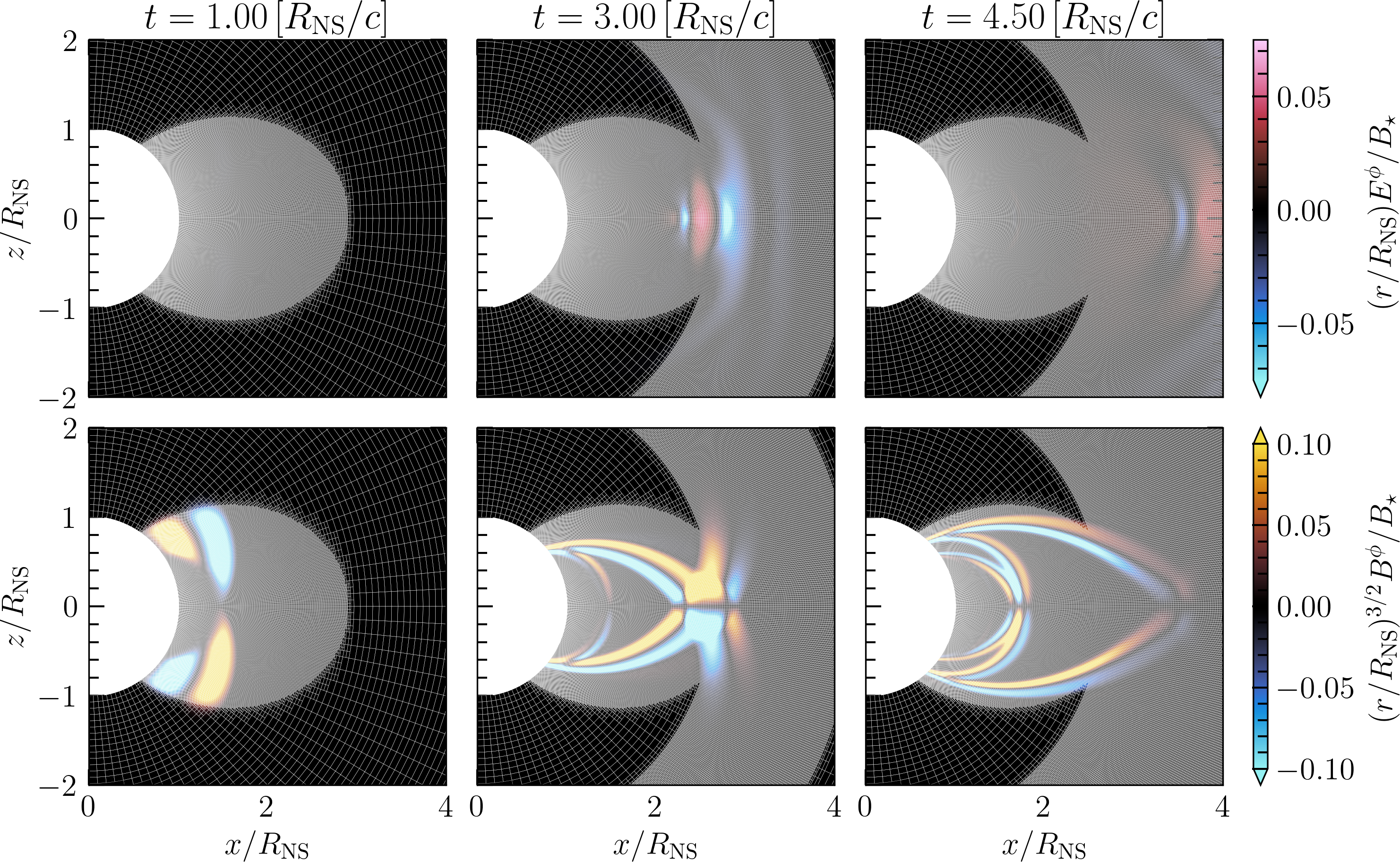}
    \caption{
    The toroidal component of the magnetic and ideal electric field, $B^\phi$ and $E^\phi$, for an MHD monster shock initiated through the launching of A waves from the stellar boundary. The outline of mesh blocks is shown in white with, each block containing 16 radial cells and 8 angular cells. The plotted simulation has domain $r/R_{\rm{NS}}\in[1,10]$ and $\theta\in[0,\pi]$ with a base grid of $2048\times512$ cells and 3 levels of mesh refinement for a max effective resolution of $16384\times4096$.} 
    \label{fig:amr2}
\end{figure*}

For the A waves launched through twisting the neutron star crust the resolution is maximized from the outermost field line inwards. That is, the grid is forced to the maximum AMR level for $ \sin^2\theta /r \geq 1/R_{\rm{max}}$ where
$R_{\rm{max}} =  R_{\rm{NS}}/\sin^2 \left( \theta_0 - \Delta \right)$, $\theta_0$ is the center of the twist, and $\Delta$ is the angular half-width. Here we have made use of the relation of $ \sin^2\theta = r/R$ for a dipolar field line, where $R$ is the radial position of the field line on the equator. The A waves will produce fast modes through mode conversion; we are particularly interested in the case where two large amplitude A waves collide at the equator and produce a fast wave with a substantial amplitude. We track these produced fast modes with AMR by initiating an outgoing AMR wave region, similar to the case with the fast mode launched from the stellar surface, but now assuming the A waves travel along the field line at the speed of light and produce fast waves starting at the equator. Let $\tau = S/c$ be the time for the wave to propagate to the equator along the innermost twisted field line, then
\begin{align}
    S & = \int_{\theta_0+\Delta}^{\pi/2} \sqrt{\left( \frac{\partial r}{\partial \theta} \right)^2 + r^2} \, d\theta\\
    & = R_{\rm{max}}\int_{\theta_0+\Delta}^{\pi/2} \sqrt{ \sin^2(2\theta)+ \sin^4(\theta) } \, d\theta\\
    & = R_{\rm{max}} \left[ \frac{1}{12} \left(
    \begin{aligned}[t]
        & -\sqrt{3} \left( \ln 2 - 2 \ln\left( \sqrt{6} \cos(\theta_0+\Delta) + \sqrt{5 + 3 \cos(2(\theta_0+\Delta))} \right) \right) \\
        & + 6 \cot(\theta_0+\Delta) \sqrt{\sin^4(\theta_0+\Delta) + \sin^2(2(\theta_0+\Delta))}
    \end{aligned}
    \right) \right].
\end{align}
For $t > \tau - 0.25 R_{\rm{NS}}/c$, we force the grid to the maximum AMR level for $ r_ - < r < r_+$, where
\begin{align}
    r_+ & = R_{\rm{max}} + c(t-\tau + 0.25 R_{\rm{NS}}/c),\\
    r_- & = R_{\rm{max}} - 0.25 R_{\rm{NS}}.
\end{align}
This refinement of the field region along which the A waves propagate and the tracking of the fast waves are shown in Figure~\ref{fig:amr2}.

Simulations which launch a fast wave from the surface through a wrinkled dipolar magnetic field only apply AMR to follow the fast wave as it moves through the magnetosphere, leaving the remainder of the magnetosphere at the base resolution. 



\begin{figure*}
    \centering
    \includegraphics[width=\linewidth]{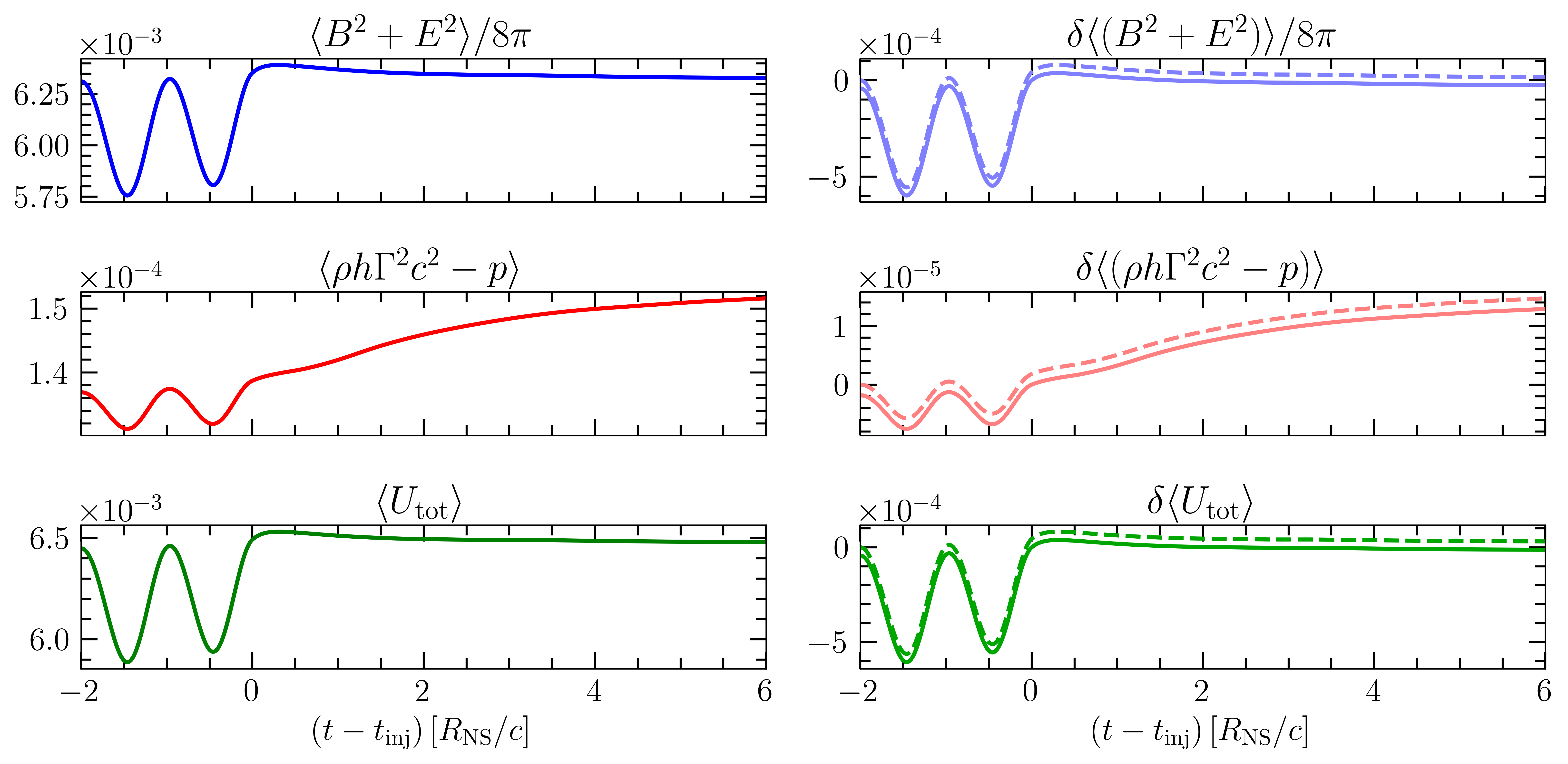}
    \cprotect\caption{The time dependence of area averaged quantities over the entire domain for simulation \Verb+sigma100+. Top row: the total and change in electromagnetic energy.
    Middle row: the total and change in thermal and kinetic energy in the plasma.
    Bottom row: the total and change in total energy.
    For all plots in the right column, the solid line shows the change from the quantity at time $t_{\rm{inj}}$, where the wave has finished injection from the boundary, and the dashed line shows the change from the quantity at $t=0$.}
    \label{fig:energy}
\end{figure*}

Throughout the simulation run time we set $T = p/\rho=0$ outside of regions with $E > 0.7 B$ or $T  > 10^{-3}$, and the velocity components parallel to the magnetic field are set to zero. Setting $\bm{v}\cdot\bm{B}=0$ sets the drift velocity to be the dominant component which is expected at larger magnetization as in the shock $(E^2 -B^2)/B^2 = 1/\Gamma_D^2 \propto 1/\sigma^2$, $v_D \rightarrow 1$ so $v_\parallel \rightarrow 0$ (\citetalias{2023ApJ...959...34B}, \citealt{2024ApJ...975..223B}). This minimizes the low magnetization effects. 
In Figure~\ref{fig:energy}, the time evolution of area averaged quantities over the entire computational domain for FMS monster shock simulation \texttt{sigma100} is shown. The wave is injected for $t-t_{\rm{inj}}<0$, during which the electromagnetic energy changes due to the polarization of the wave, since one half of the wave acts to cancel the background field and the other adds with the background. The total thermal and kinetic plasma energy changes because the magnetization $\sigma$ is held constant at the stellar boundary, thus when the wave subtracts from the dipolar background the density goes down, and when the wave adds to the background the density goes up. Additionally, a small amount of kinetic energy is injected due to the wave being injected through the drift velocity. 
Approximately $\sim0.4\%$ of the total energy is lost during the run time of the simulation, through the floors described above, numerical dissipation, and the entropy evolution described in Appendix~\ref{app:num_methods}. At the resolutions presented in the main text, numerical dissipation will be negligible \citep{2025PhRvD.112f3046G}.


\begin{figure}
    \centering
    \includegraphics[width=0.4\linewidth]{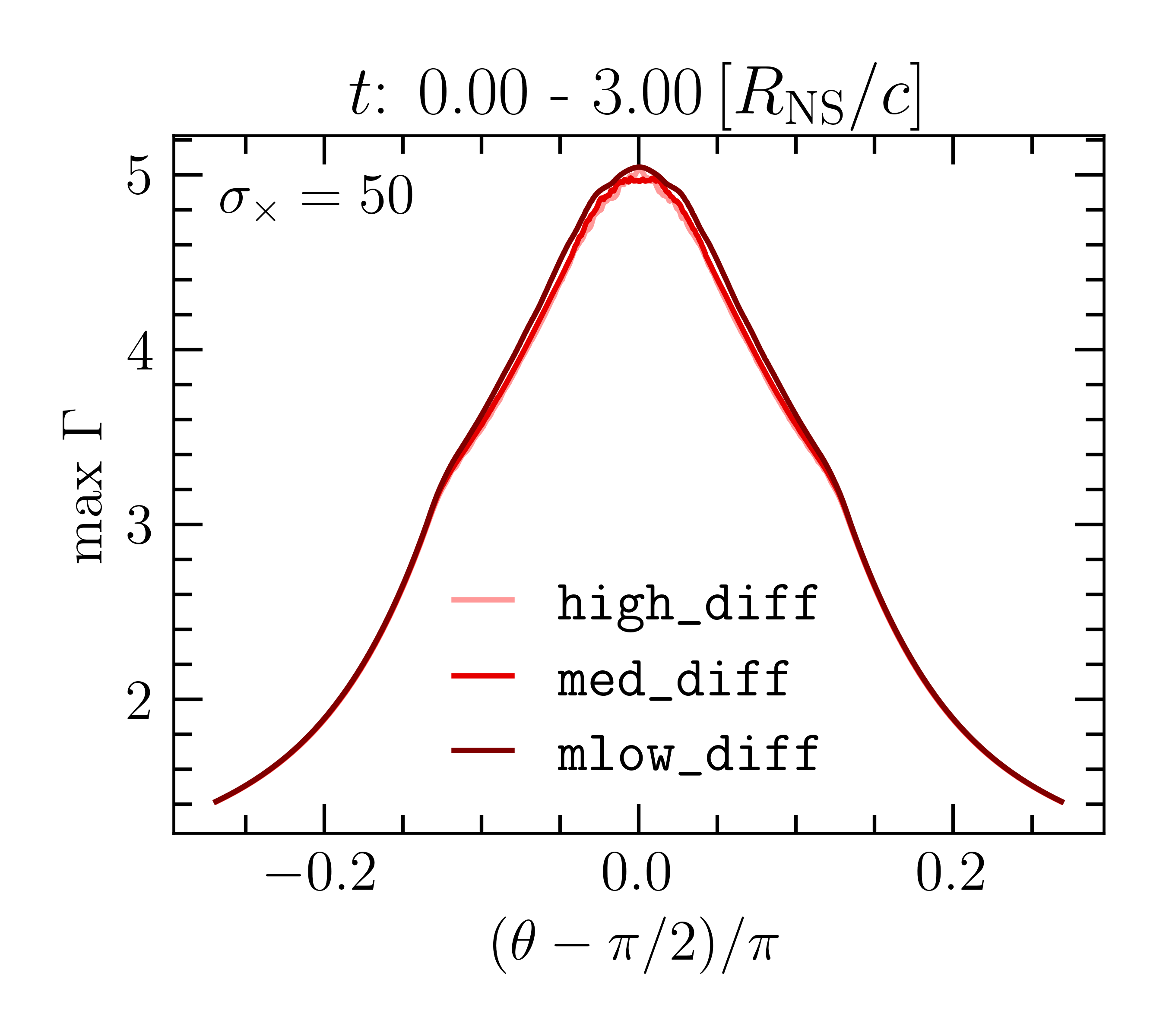}
    \includegraphics[width=0.4\linewidth]{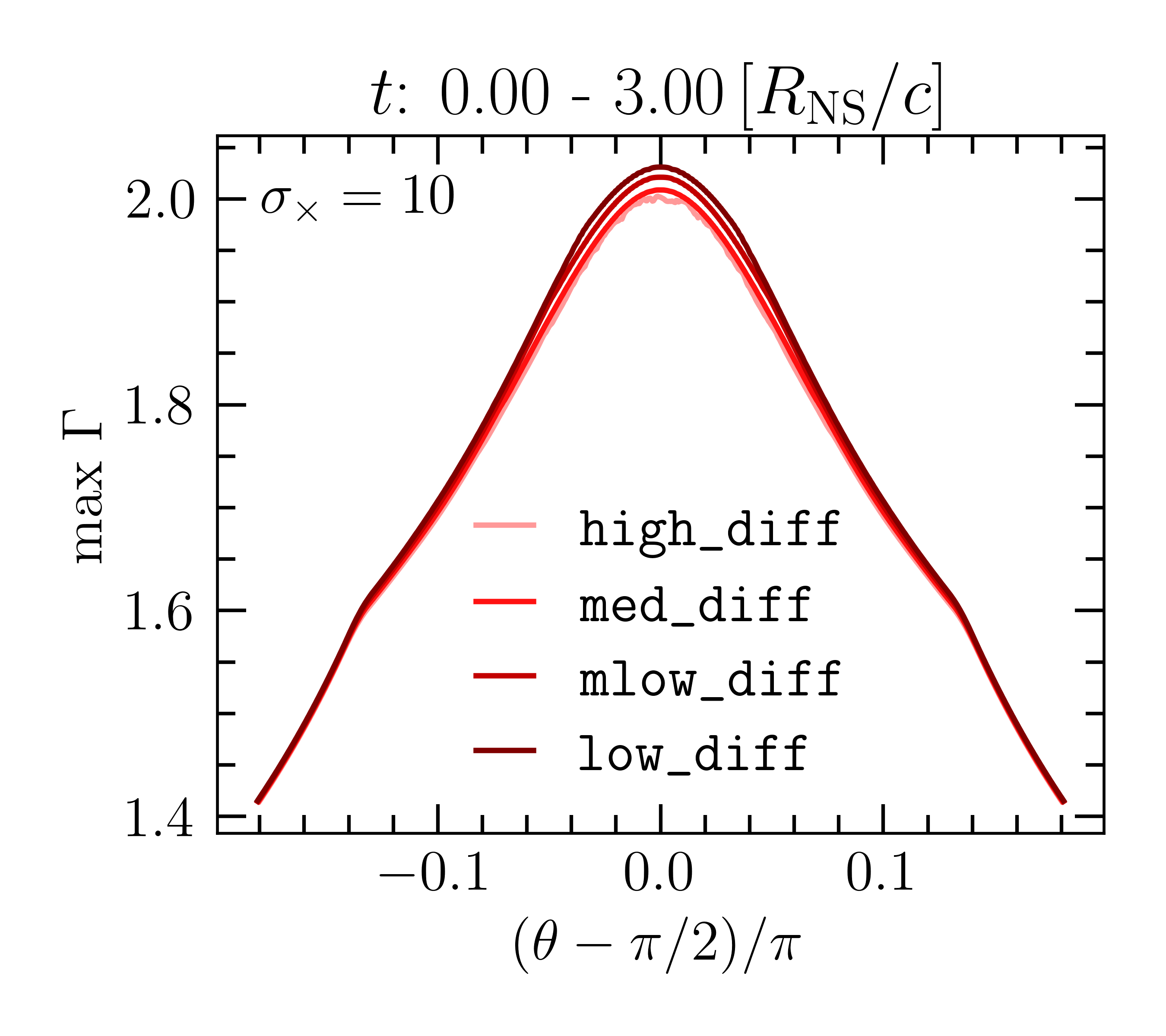}
    \caption{
    The max Lorentz factor as a function of poloidal angle 
    for simulations with $\sigma_\times = 50$ (left) and $\sigma_\times=10$ (right). 
    The value of the diffuse flux coefficient, $D$, is as follows: 
    \texttt{low\_diff}, $D=1$; \texttt{mlow\_diff}, $D=2$; 
    \texttt{med\_diff}, $D=5$; and \texttt{high\_diff}, $D=10$.
    }
    \label{fig:diff}
\end{figure}

To maintain numerical stability at large magnetization, additional numerical diffusion is required,; for the Total Variation Diminishing Lax-Friedrichs (TVDLF) method the diffuse flux part has a coefficient, $D=1$, by default. To maintain stability for $\sigma_\times = 100$, we set $D=10$, ten times the default. The additional numerical diffusion mildly reduces the peak Lorentz factor of the shock, as shown in Figure~\ref{fig:diff}.
This reduction is consistent across magnetizations and reduces the $y$-intercept of the linear dependence of $\Gamma$ on $\sigma_\times$. We elect to maintain a constant increased diffusion, $D=10$, across simulations where the magnetization is varied, even in low magnetization configurations which do not require it for numerical stability. This choice of $D$ is made by determining the lowest value which maintains numerical stability for $\sigma_\times=100$.  Simulations which vary the FMS frequency or amplitude of the wave are performed with $\sigma_\times = 50$ which requires less additional diffusion to maintain stability, and hence we choose $D=5$ for these simulations. Simulations which vary the number of FMS wavelengths launched are performed with $\sigma_\times = 50$ but set $D=10$, as the shocks of subsequent wavelengths become increasingly numerically challenging. All Alfv\'{e}n mode conversion and wrinkled background simulations set either a uniform $\sigma_{\rm bg}= 100$ or $\sigma_\times=100$, therefore we set $D=10$ to maintain numerical stability at large magnetization.


\begin{figure}
    \centering
    \includegraphics[width=0.49\linewidth]{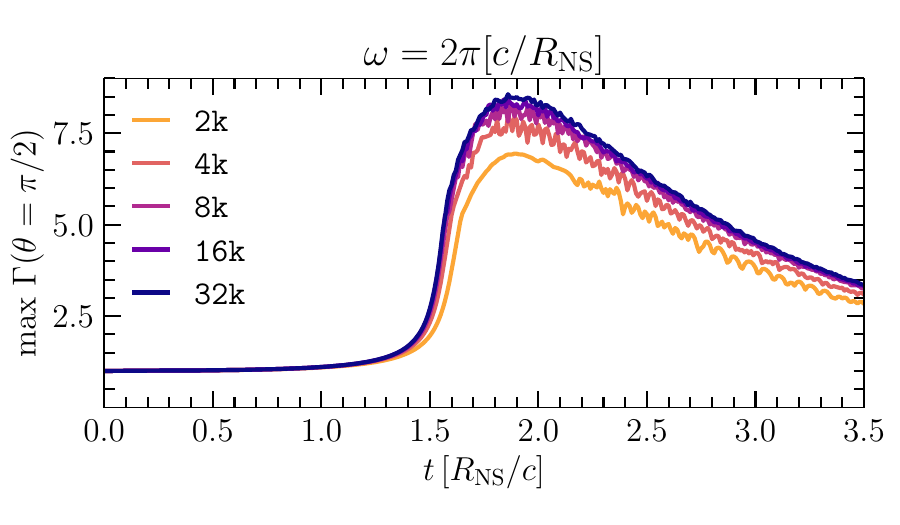}
    \includegraphics[width=0.49\linewidth]{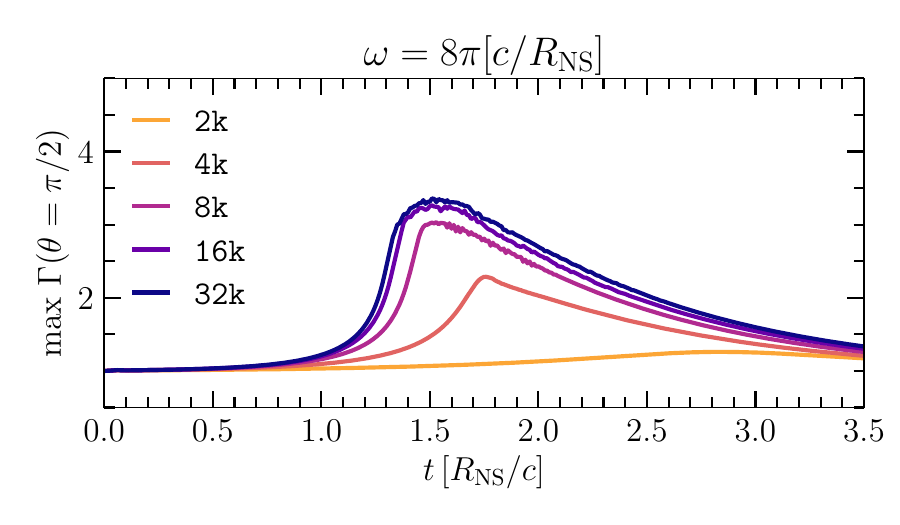}
    \caption{
    Convergence tests for
    the maximum on-equator Lorentz factor of a single isotropic FMS wave for $\sigma_\times = 100$ with $\omega = 2\pi \, [c/R_{\rm NS}]$ (left) and $\omega = 8\pi \, [c/R_{\rm NS}]$ (right). The number of radial cells, $N_r \in[2048, 4096, 8192, 16384, 32768]$, for $r/R_{\rm NS}\in[1,10]$ is stated in the legend, there are eight times fewer angular cells for $\theta\in[0,\pi]$. 
    }
    \label{fig:con_test}
\end{figure}

\section{Numerical Convergence}\label{app:convergence}


We assess convergence with the isotropic FMS wave configuration by gradually lowering the resolution from double that adopted in the main text and identifying the point at which the equatorial Lorentz factor begins to diverge for $\sigma_\times=100$ and $E_{\rm w}/B_\star=0.1$, with $\omega = 2\pi \, [c/R_{\rm NS}]$ and $\omega = 8\pi \, [c/R_{\rm NS}]$. As shown in Figure~\ref{fig:con_test}, for $\omega = 2\pi \, [c/R_{\rm NS}]$ the equatorial shock is well resolved with 16,384 radial cells and moderately resolved with 8,192 radial cells, while for $\omega = 8\pi \, [c/R_{\rm NS}]$ the equatorial shock is well resolved with 16,384 radial cells. Notice that $\omega = 8\pi \, [c/R_{\rm NS}]$ diverges significantly at lower resolutions ($N_r \leq 4096$), indicating that at such resolutions the wave itself is unresolved. Once the wave is resolved it appears the resolution required to resolve the shock is similar between $\omega = 2\pi \, [c/R_{\rm NS}]$ and $\omega = 8\pi \, [c/R_{\rm NS}]$. From this we conclude the simulations presented in the main body of the text are resolved. The lowest radial resolution simulations are those with a wrinkled background, for which we use $N_r = 8192$. This resolution is sufficient to moderately resolve the $\omega = 2\pi \, [c/R_{\rm NS}]$ FMS wave launched into the perturbed background. In these runs we chose the lower radial resolution to keep computational costs manageable, allowing us to increase the angular resolution to $N_\theta = 8192$ in order to better capture the standing waves and shock front fragmentation.

\begin{figure*}
    \centering 
    \gridline{
      \fig{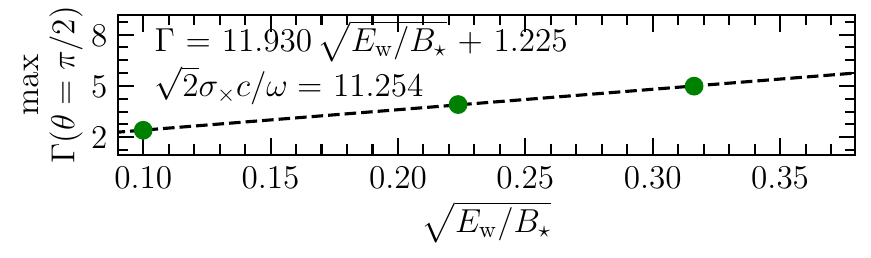}{0.49\textwidth}{}
      \fig{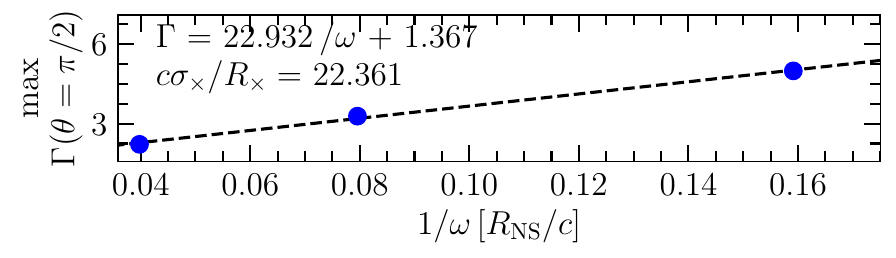}{0.49\textwidth}{}
    }
    \vspace{-25pt}
    \gridline{
      \fig{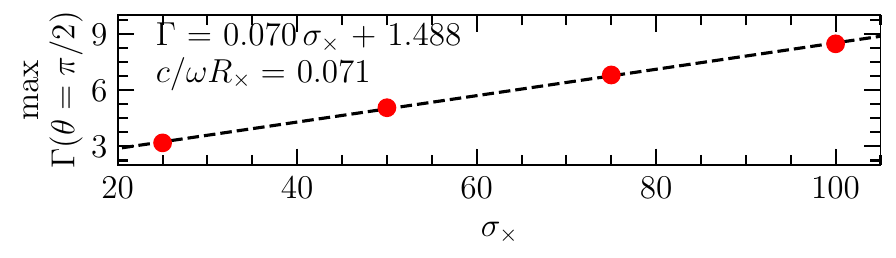}{0.49\textwidth}{}
    }
\vspace{-25pt}
\caption{
The maximum Lorentz factor of the equatorial monster shock as a function of the wave amplitude $E_{\rm{w}}$ (top left); the wave frequency $\omega$ (top right); and the magnetization $\sigma_\times$ (bottom). The black dashed line is a linear fit, which produces the expected slope and a non-zero $y$-intercept. On each plot the linear fit is displayed along with the expected slope from Equation~\eqref{eq:lfac_equator}. Details of the isotropic FMS wave simulations can be found in Table~\ref{tab:sims}. All values are taken on the equator, $z/R_{\times}=0$.
}
    \label{fig:lfac_summary2}
\end{figure*}

\section{Linear Fitting}\label{app:fitting}

To verify the scalings predicted by Equation~\eqref{eq:lfac_equator}, we perform linear fits to the equatorial upstream Lorentz factor of an isotropic FMS wave without fixing the slope. As shown in Figure~\ref{fig:lfac_summary2}, the Lorentz factor recovers the expected linear dependence on wave amplitude, wave frequency, and background magnetization, and the fitted slopes closely match the analytic predictions.

\section{Cylindrical Geometry}\label{app:cyl}

\begin{figure*}
    \centering
    \includegraphics[width=\textwidth]{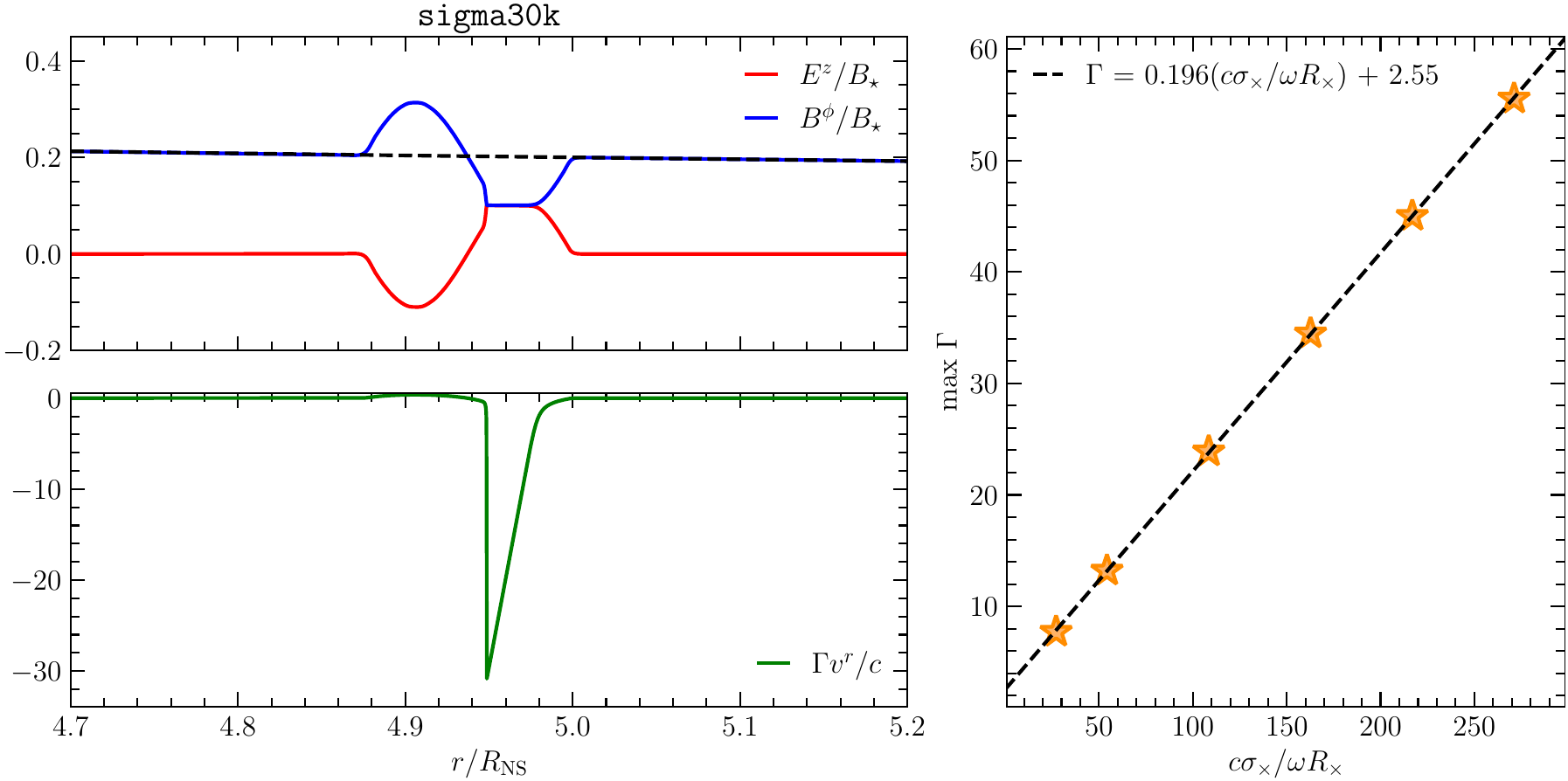}
    \caption{
    The quasi-1D cylindrical monster shock for $\sigma_\times \in [5,10,20,30,40,50]\times 10^3$ with $\lambda/R_\times = 0.034$.
    Left: A snapshot of \texttt{sigma30k} at $t=4.0 \, R_{\rm NS}/c$, showing the background and wave magnetic field $B^\phi$, the wave electric field $E^z$, and the radial four velocity $\Gamma v^r$. A dashed black line shows the $(R_{\rm NS}/r)$ scaling of the background magnetic field. 
    Right: The maximum Lorentz factor of the cylindrical monster shock as a function $c \sigma_\times/\omega R_\times$. The black dashed line is a linear function with a fitted slope and intercept. 
    }\label{fig:appcyl_scaling}
\end{figure*}

In this Section~we consider the launching of an isotropic FMS wave in cylindrical coordinates $(r,\phi,z)$ \citep{2022arXiv221013506C}.
The background magnetic field is purely in the $\phi$-direction, $\bm{B}_{\rm bg} = B_\star (R_{\rm NS}/r) \hat{\phi}$, and the wave is launched through a boundary condition on the electric field $\bm{E}(r=R_{\rm NS})=\sqrt{R_{\rm NS}/r} E_{\rm w} \sin(\omega t) \hat{z}$. The FMS wave grows relative to the background as $B_{\rm w}/B_{\rm bg} \propto \sqrt{r}$ and therefore the non-linearity radius is located at 
\begin{equation}
    R_\times = R_{\rm NS} \left( \frac{B_\star}{2E_{\rm w}} \right)^2.
\end{equation}
We perform a suite of quasi-1D cylindrical simulations that enable us to probe the upstream Lorentz factor scaling at larger magnetizations. 
All simulations presented in this Section~have $\lambda/R_\times = 0.034$, with $\sigma_\times \in [5,10,20,30,40,50]\times 10^3$, and have a uniform resolution of $(131072,16)$ cells in $(r,\phi)$ with $r/R_{\rm{NS}}\in[1,10]$ and $\phi = [-d\phi, d\phi]$ with $d\phi=0.01$.
This increase in magnetization is possible because quasi-1D simulations are far cheaper and can therefore be run at much higher radial resolution, and because the numerical difficulties that appear off the equator in 2D do not arise here, as the shock is strictly perpendicular.
All cylindrical simulations set the magnetization as
\begin{equation}
\sigma_{\rm bg}(r) = \frac{B_{\rm{bg}}^2}{4\pi \rho_{\rm{bg}}c^2 } =
\sigma_\times
\begin{cases}
1, & r < R_\times, \\[6pt]
\left( \dfrac{R_\times}{r} \right)^2, & r \ge R_\times,
\end{cases}
\end{equation}
and use the same flooring scheme detailed in Appendix~\ref{app:amr_flooring}, but without the need for any additional numerical diffusion. In the left panel of Figure~\ref{fig:appcyl_scaling}, a shock forms through the same mechanism as shown in dipolar geometry, the wave field is polarized such that $E^2 \rightarrow B^2$ and half of the wave is shaved off, forming a shock with huge upstream Lorentz factor. 
As shown in the right panel of Figure~\ref{fig:appcyl_scaling}, the Lorentz factor produced through the monster shock mechanism in cylindrical geometry still demonstrates a linear scaling in $c\sigma_\times/\omega R_\times$, albeit with a smaller than unity slope (expected slope derived below) and a different constant offset.

Here we repeat the analysis of \citetalias{2023ApJ...959...34B} but for cylindrical geometry, demonstrating the linear scaling of the maximum upstream Lorentz factor. 
The energy evolution equation for the MHD flow in a (short) wave may be written as \citepalias{2023ApJ...959...34B}
\begin{equation}
    (\partial_t \delta \mathcal{E})_\xi = - \dot{\mathcal{N}}_\xi m c^2 \delta\Gamma, \label{eq:energyeqn}
\end{equation}
where $\xi = t - r/c$, $\dot{\mathcal{N}}_\xi$ is the number of particles crossing a given $\xi$ per unit time, $\delta\mathcal{E}$ is the wave energy contained in interval $\delta \xi$, and $\delta\Gamma$ is the change of $\Gamma$ of the MHD flow as it crosses $\delta\xi$. In cylindrical geometry, the wave energy per unit length measured along the axis and the particle flux per unit length are
\begin{equation}
    \delta\mathcal{E} = 2 \pi rc\left( \frac{E^2}{4\pi}\right)\delta\xi, \qquad \dot{\mathcal{N}}_\xi = 2\pi rc n_{\rm bg}.
\end{equation}
The change $\delta\Gamma$ of a fluid element can be expressed in terms of $\delta\xi$ using the derivative along the worldline where $d\xi = dt - dr/c = (1-\beta)dt$,
\begin{equation}
    \frac{\delta\Gamma}{\delta\xi} = \frac{d x^\alpha}{d\xi}\partial_\alpha \Gamma = \frac{(\partial_t\Gamma)_\xi}{1-\beta} + \partial_\xi \Gamma,
\end{equation}
note that the repeated indices runs over the coordinates $(t,\xi,\phi,z)$ not $(t,r,\phi,z)$.
Equation~\eqref{eq:energyeqn} then becomes 
\begin{equation}
    \partial_t \left(r \frac{E^2}{4\pi} \right)_\xi  = - r \rho_{\rm bg} c^2 \left[ \frac{(\partial_t\Gamma)_\xi}{1-\beta} + \partial_\xi \Gamma \right]. \label{eq:energyeqn_cyl}
\end{equation}
It differs from the spherical (dipolar $B_{\rm bg}$) case only by changing the geometric factor $r$ in the equation: $r \leftrightarrow r^2$.
Following the method used in \citetalias{2023ApJ...959...34B} one can rewrite Equation~\eqref{eq:energyeqn_cyl} as a differential equation for $\kappa(t, \xi)$ where $\kappa \equiv \Gamma(1 + \beta)$. First, express the wave electric field in terms of $B_{\rm bg}$ and $\kappa$,
\begin{equation}
    E = \beta B = \frac{\beta B_{\rm bg}}{1-\beta} = \frac{\kappa^2 - 1}{2} B_{\rm bg}.
\end{equation}
Here, we used $B = B_{\rm bg}/(1 - \beta)$, which describes compression of $B$ in the short MHD wave. The left-hand side of Equation~\eqref{eq:energyeqn_cyl} may be expressed as
\begin{equation}
    \partial_t (r E^2)_\xi = -c E^2 + 2rE B_{\rm bg}\kappa(\partial_t \kappa)_\xi,
\end{equation}
where we used $(\partial_t B_{\rm bg})_\xi = (\partial_t r)_\xi dB_{\rm bg}/dr = - c B_{\rm bg}/r$, since $B_{\rm bg} \propto r^{-1}$ in the cylindrical problem. The right-hand side of Equation~\eqref{eq:energyeqn_cyl} may be expressed in terms of $\kappa$ using $1 - \beta = 2/(\kappa^2 + 1)$,
$\Gamma = (\kappa^2 + 1)/2\kappa$ and $d\Gamma/d\kappa = (\kappa^2 - 1)/2\kappa^2$. Then, Equation~\eqref{eq:energyeqn_cyl} becomes
\begin{equation}
-c E^2 + 2rE B_{\rm bg}\kappa(\partial_t \kappa)_\xi
= - 4\pi r \rho_{\rm bg} c^2 \left[ \frac{1}{2}(\kappa^2 + 1)(\partial_t \kappa)_\xi +\partial_\xi \kappa \right] \frac{(\kappa^2 -1)}{2\kappa^2}.
\end{equation}
This gives an  equation for $\kappa(t,\xi)$,
\begin{equation}
    \left(  2 \sigma_{\rm bg} \kappa^3 + \frac{1+\kappa^2}{2}  \right) \partial_t \kappa + \partial_\xi \kappa = \frac{c \sigma_{\rm bg}}{2r} \kappa^2 (\kappa^2 - 1),
\end{equation}
which can be further simplified by neglecting the $\kappa^2/2$ term in the limit $1/\sigma_{\rm bg} \kappa \ll 1$,
\begin{equation}
    \left(  2 \sigma_{\rm bg} \kappa^3 + \frac{1}{2}  \right) \partial_t \kappa + \partial_\xi \kappa = \frac{c \sigma_{\rm bg}}{2r} \kappa^2 (\kappa^2 - 1). \label{eq:kappa_cyl}
\end{equation}
Equation~\eqref{eq:kappa_cyl} differs from the corresponding equation in spherical (dipolar field) geometry (Equation~(41) in \citetalias{2023ApJ...959...34B}) by the factor of $1/4$ on the right-hand side. So, the acceleration of the MHD flow along the plateau may be expected to be approximately 4 times slower in the cylindrical problem. The rate of change of $\kappa$ along the $C^+$ characteristics is
\begin{equation}
   \left. \frac{d\kappa}{d\xi}\right|_{C^+} = \frac{c \sigma_{\rm bg}}{2r} \kappa^2 (\kappa^2 - 1), \label{eq:Cplus}
\end{equation}
where the derivative is taken along the $C^+$ curve defined by \citepalias{2023ApJ...959...34B} 
\begin{equation}
    \frac{d\xi_+}{dt} = \frac{2}{4\sigma_{\rm bg} \kappa^3 +1}.
\end{equation}
In particular, on the plateau $\kappa \ll 1$ and $\Gamma \approx (2\kappa)^{-1}$, and Equation~\eqref{eq:Cplus} gives
\begin{equation}
    \left. \frac{d\Gamma}{d\xi}\right|_{C^+} = \frac{c\sigma_{\rm bg}}{4r} \qquad (\kappa\ll 1).
    \label{eq:dgammadxi}
\end{equation}
The resulting $\Gamma$ flow entering the monster shock is given by
\begin{equation}
    \Gamma = \frac{\sigma_{\rm bg} W_{\rm p}}{4r},
\end{equation}
where $W_{\rm p}$ is the width of the plateau. For a sine wave the boundaries of the plateau are defined by the condition $\sin(\omega \xi) = -B_{\rm bg}/2E_0 = (R_\times/r)^{1/2}$, taking into account that $B_{\rm bg} \propto r^{-1}$ and the wave amplitude $E_0 \propto r^{-1/2}$ in the cylindrical problem, this gives
\begin{equation}
    W_{\rm p}(r) = \frac{c}{\omega} \left[ \pi - 2 \arcsin \left( \frac{R_\times}{r} \right)^{1/2} \right]. \label{eq:Wp_cyl}
\end{equation}
Therefore, we expect that $\Gamma \propto c\sigma_{\rm \times}/\omega R_{\times}$ just as in the spherical problem but with an additional factor of $1/4$ and a modified geometrical factor given by the peak of $\sigma_{\rm bg} W_{\rm p}$.

\begin{figure*}
\centering
\gridline{
  \fig{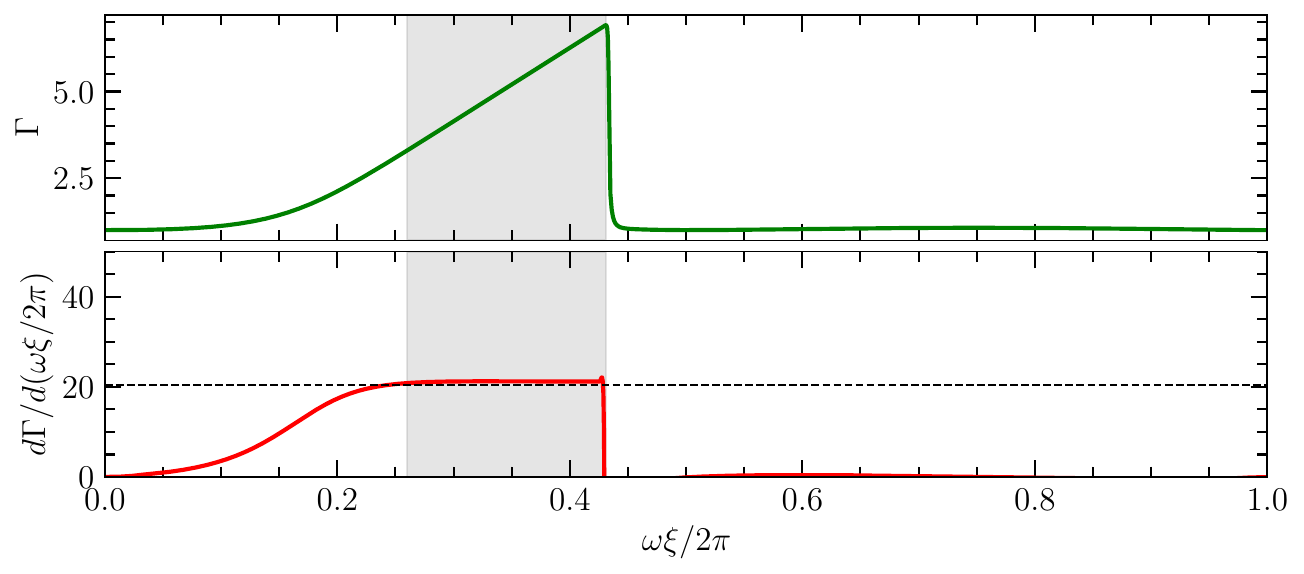}{0.85\textwidth}{(a) \texttt{sigma5k}}
}
\gridline{
  \fig{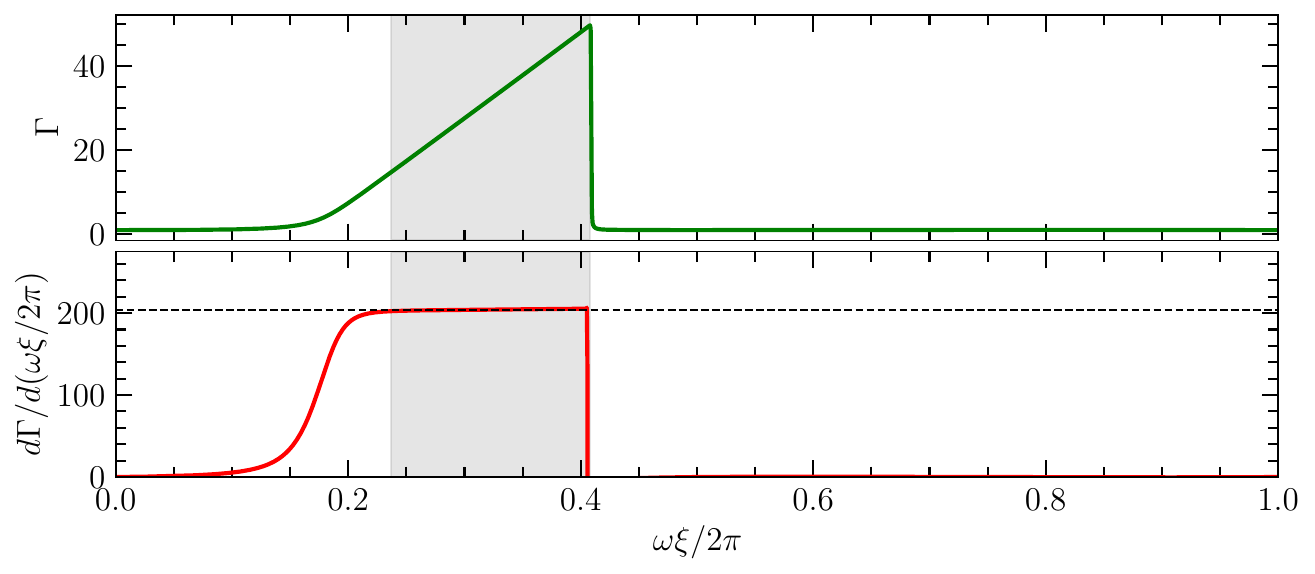}{0.85\textwidth}{(b) \texttt{sigma50k}}
}
\caption{
    The Lorentz factor (top) and the derivative of the Lorentz factor with respect to the light cone coordinate $\xi$ (bottom) of the quasi-1D cylindrical monster shock in simulations \texttt{sigma5k} and \texttt{sigma50k} at $t = 4.0 R_{\rm NS}/c$. The analytical expectation (Equation~\eqref{eq:dgammadxi}) for the derivative is plotted as a black dashed line in the bottom panel, and the analytical plateau width (Equation~\eqref{eq:Wp_cyl}) is shown by the gray region in both panels.
}
\label{fig:sigma5000}
\end{figure*}

For the choice of magnetization profile made for the cylindrical simulations in this section,
\begin{equation}
    \rm{max}\, \Gamma \approx  0.5 \, \frac{1}{4}\left( \frac{c \sigma_\times}{ \omega R_{\times}}\right),
\end{equation}
with peak at $r/R_\times \approx 1.17$. 
As in the spherical case, the measured slope does not fully agree with the analytical prediction. In the cylindrical simulations we find a slope of $\sim 0.2$, compared to the expected value of $\sim 0.125$. For reference, in the spherical case we measure a slope of $\sim 1$, whereas the analytical prediction is $\sim 0.82$. Owing to the numerical robustness of the quasi-1D setup, we can probe this discrepancy using the simulations \texttt{sigma5k} and \texttt{sigma50k}, with $\sigma_\times = 5\times10^3$ and $\sigma_\times = 50
\times10^3$ both with $\lambda/R_\times = 0.034$. As shown in Figure~\ref{fig:sigma5000}, the slope ($d\Gamma/d\xi$) agrees well with the analytical prediction. However, the plateau is wider than expected and exhibits modest pre-acceleration (relative to the maximum Lorentz factor), leading to a larger-than-predicted maximum upstream Lorentz factor. This arises because the plateau width is estimated assuming force-free propagation with $E = B = B_{\rm bg}/2$, which neglects the upstream region where the Lorentz factor is still well below its maximum.

    

\clearpage
\bibliography{ref}

\end{document}